\documentclass[11pt,a4paper]{article}
\usepackage{jheppub}
\usepackage[utf8]{inputenc}
\usepackage{fancybox,float,comment}
\usepackage[dvipsnames]{xcolor}
\usepackage{units}
\usepackage{subcaption}
\usepackage{multirow,array,arydshln}
\usepackage{yfonts}
\usepackage{tensor}
\usepackage{paralist}
\usepackage{tensor,mathrsfs}
\usepackage{ulem}
\usepackage{mathtools}
\usepackage{amsmath}
\usepackage{amssymb}
\usepackage{setspace}
\usepackage{amsfonts}
\usepackage{color}
\usepackage{threeparttable}
\usepackage{cancel}
\usepackage{tikz}
\usetikzlibrary{decorations.pathmorphing}
\usepackage{graphicx}
\usepackage[raggedright]{sidecap}
\usepackage[utf8]{inputenc}
\usepackage{multirow}
\usepackage{hhline}
\usepackage{colortbl}
\usepackage{diagbox} 
\usepackage{makecell}
\usepackage[]{hyperref}
\usepackage{rotating}
\usepackage{caption}
\usepackage{float}

\newcommand{\qn}{{\mathfrak{q}}}
\newcommand{\wn}{{\mathfrak{w}}}

\def\sec#1{section \;\ref{#1}}

\begin{document}
%

\title{Constraints on quasinormal modes and bounds for critical points from pole-skipping}
\author[a]{Navid Abbasi}
\author[b]{and Matthias Kaminski}
\affiliation[a]{School of Nuclear Science and Technology, Lanzhou University, 222 South Tianshui Road, Lanzhou 730000, China }
\affiliation[b]{Department of Physics and Astronomy, University of Alabama, Tuscaloosa, AL 35487, USA}

\emailAdd{abbasi@lzu.edu.cn}
\emailAdd{mski@ua.edu}

\date{\today}

\abstract{
We consider a holographic thermal state and perturb it by a scalar operator whose associated real-time Green's function has only gapped poles.  
These gapped poles correspond to the non-hydrodynamic quasinormal modes of a massive scalar perturbation around a Schwarzschild black brane. 
Relations between pole-skipping points, critical points and quasinormal modes in general emerge when the mass of the scalar and hence the dual operator dimension is varied. 
First, this novel analysis reveals a relation  between the location of a mode in the infinite tower of quasinormal modes and the number of pole-skipping points constraining its dispersion relation at imaginary momenta. 
Second, for the first time, we consider the radii of convergence of the derivative expansions about the gapped quasinormal modes. These convergence radii turn out to be bounded from above by the set of all pole-skipping points. 
Furthermore, a transition between two distinct classes of critical points occurs at a particular value for the conformal dimension, implying close relations between critical points and pole-skipping points in one of those two classes. 
We show numerically that all of our results are also true for gapped modes of vector and tensor operators. 
}

\maketitle
\section{Introduction}
%
Stability and causality of distinct formulations of hydrodynamics have been discussed over several decades by now~\cite{Hiscock:1985zz} and are still under debate~\cite{Kovtun:2019hdm,Hoult:2020eho,Taghinavaz:2020axp,Bemfica:2017wps,Bemfica:2019knx,Bemfica:2020zjp}. One central question is what the convergence radius of the hydrodynamic derivative expansion is when written in momentum space (after a Fourier transformation in a translation-invariant state).\footnote{The convergence of the hydrodynamic derivative expansion in position space has been recently discussed again in~\cite{Heller:2013fn,Heller:2015dha}, and the relation between position space and momentum space convergence has been discussed in the regime linearized in the hydrodynamic fields~\cite{Heller:2020uuy}.}  
This question has recently been addressed by considering the analytic continuation of the hydrodynamic derivative expansion to complex frequencies as functions of complex momenta~\cite{Withers:2018srf,Grozdanov:2019kge}. There it was shown that, at least in systems with a holographic dual, the radius of convergence is determined by the magnitude of the momentum associated with the {\it critical points of the spectral curve}, those which are the closest to the origin, which will be defined and discussed in section~\ref{sec:gapped}. Hydrodynamic (eigen)modes can be understood as those momenta and frequencies satisfying the spectral curve equation. 

Hydrodynamic (eigen)modes are relevant for the description of systems out of but close to equilibrium. Hydrodynamic perturbations are restricted to have low energies, which translates to small frequencies and small momenta. They describe the evolution of conserved charges like energy density and electric charge over large distances in space and over large time intervals. Hence, the relevance of such an expansion of gapless (eigen)modes about zero frequency and zero momentum is immediately clear.

In this work, we consider gapped (eigen)modes and the radius of convergence of derivative expansions about those gapped mode locations. There are several physical motivations for this study. First, most systems in Nature only contain gapped (eigen)modes. In that case, the lowest-lying mode, the one with the smallest imaginary part of its frequency, dominates the time-evolution of the state. Hence, an expansion about that mode location allows a perturbative description of the time-dependence of the state. 
Examples of expansions around gapped modes are (i)~perturbations around the Fermi surface, or (ii)~perturbations around a large mass or energy scale leading to non-relativistic excitations. Second, even in systems which contain both gapless and gapped modes, the gapped modes can dominate the evolution of the state. Consider for example a system which contains a diffusion mode, i.e.~a mode with purely imaginary (quasi-eigen)frequency, $\mathfrak{w}\propto \mathfrak{q}^2$, with the momentum $\mathfrak{q}$ directed along one of the spatial coordinates. At a particular value of the momentum $\mathfrak{q}$ the lowest non-hydrodynamic mode has a smaller imaginary part than the hydrodynamic mode, thus becoming the dominant QNM for larger momenta. Note that this transition from a regime of hydrodynamic mode dominance to non-hydrodynamic mode dominance has been pointed out  before~\cite{Amado:2008ji,Kaminski:2009ce}. 
The same change in dominance of QNMs can happen when certain parameters of the system are changed, for example the charge of a Reissner-Nordstr\"om black brane~\cite{Janiszewski:2015ura}. Lastly, our results establish relations between {\it pole-skipping points}~\cite{Blake:2017ris}, the {\it critical points} and the general locations of gapped QNMs which have not been established for the gapped modes until now. 

In thermal equilibrium, the spectrum of quasinormal modes~(QNMs) is holographically dual to the poles in the retarded Green's function of a given operator. The values of the QNMs determine how fast perturbations of that operator thermalize. For operators without gapless excitations, the associated spectrum contains a tower of infinitely many gapped QNMs.  
The  lower a mode is in the spectrum, the smaller is the imaginary part of its frequency and consequently it will be swallowed later by the medium. 

Recently, it has been shown that the pole-skipping phenomenon, at least in holographic systems, forces the response functions associated with operators in the theory to satisfy an infinite number of constraints \cite{Blake:2019otz,Grozdanov:2019uhi} \footnote{The pole-skipping phenomenon was first observed as the indication for the hydrodynamic origin of quantum chaotic behavior~\cite{Grozdanov:2017ajz,Blake:2017ris,Blake:2018leo,Grozdanov:2018kkt}. Then it was investigated in many other examples and with distinct motivations~\cite{Wu:2019esr,Ahn:2019rnq,Li:2019bgc,Ceplak:2019ymw,Das:2019tga,Abbasi:2019rhy,Liu:2020yaf,Ahn:2020bks,Moitra:2020dal,Grozdanov:2019uhi,Abbasi:2020ykq,Jansen:2020hfd,Grozdanov:2020koi,Ramirez:2020qer,Choi:2020tdj,Ahn:2020baf,Natsuume:2020snz,Arean:2020eus,Kim:2020url,Sil:2020jhr}.}.    
These constraints are discrete in the sense that they occur at discrete pole-skipping frequencies. Recalling that the spectrum of QNMs is discrete, too, one may be tempted to look for a possible relation between these two sets of discrete data. That task is what the first part of this paper is devoted to. In section~\ref{sec:dispersion_constraint}, we will show that pole-skipping points constrain QNMs in a way more severe than what has been known so far.

We start with computing the spectrum of gapped QNMs associated with a massive scalar field in the bulk of AdS$_5$. We will mostly work at imaginary momenta, since the  pole-skipping
points in the dual theory exclusively occur at such values.   As one might expect, we find that all pole-skipping points are lying on the dispersion relations of QNMs. But the non-trivial part of our result is that we see that \textit{the dispersion relation of the $n^{th}$ QNM is constrained by pole-skipping points with the frequencies  $\omega= m(i 2\pi T)$ where $m\ge n$.} It simply means that the lowest QNM is constrained by  pole-skipping points at all possible frequencies and consequently is the most constrained one. However, by every one level going higher in the spectrum, the  number of pole-skipping frequencies constraining the dispersion relation decreases by one. 

In order to get more insight into these relations among QNMs, in section \ref{sec:gapped}, we study the analytic structure of the spectral curve, see~\cite{Withers:2018srf,Grozdanov:2019kge} and \cite{Grozdanov:2020koi}. 
We discuss to what extent the derivative expansion is reliable for probing the dispersion relation of a given gapped QNM. 
Our focus lies on computing  the radius of convergence of the derivative expansion about the {\it lowest-lying} QNM with mode number $n=1$, as well as the {\it second} QNM with mode number $n=2$. 
We perform this analysis for scalar operators of conformal operator dimensions within the range $2\le \Delta \le 8$. This is achieved by finding the critical points of the spectral curve at complex momenta. It turns out that there exists a  transition value $\Delta_{\text{t}}$ which distinguishes between two types of critical points. For the values of $\Delta$ lower than $\Delta_{\text{t}}$, the critical point 
displays a behavior which we refer to as {\it the lowest-level-degeneracy} of the two lowest-lying complexified QNMs. 
In this case, the real part of the critical frequency vanishes, i.e.~$\text{Re}\,\omega_c=0$. In other words, within the complex frequency plane the two lowest-lying QNMs collide with each other on the imaginary frequency axis, as illustrated in the upper two plots in figure~\ref{fig:Scalar_cross_before}.  
Beyond $\Delta_{\text{t}}$ however, the relevant critical point of the spectral curve  is identified with the {\it level-crossing} of QNMs at complex momentum and complex frequency, as it was described previously~\cite{Grozdanov:2019kge,Grozdanov:2019uhi,Abbasi:2020ykq,Jansen:2020hfd,Heller:2020uuy,Baggioli:2020loj}.  
This is illustrated in figure~\ref{fig:Scalar_cross_beyond}. 
Right at the transition point, $\Delta=\Delta_\text{t}$, separating these two classes of critical points, there are three critical points which simultaneously determine the radius of convergence of the derivative expansion about the lowest-lying QNM, see figure~\ref{fig:Scalar_cross_at}. 

In section~\ref{bound} we find that pole-skipping points put a strong constraint on the position of the aforementioned critical points.  
We observe that the critical point is always enclosed within the circle passing through the nearest pole-skipping point in the $\text{Im}\wn-\text{Im}\qn^2$ plane.  
In other words, the critical point and hence the radius of convergence of the derivative expansion about the lowest-lying mode is bounded form above. 
This upper bound on the radius of convergence is illustrated in figure~\ref{fig:Scalar_cross}.

Moreover, we have shown that all the observations we state for scalar operators do also hold for vector and tensor operators, as discussed in section~\ref{sec:vectorTensor}.  
In order to relate our numerical results to a simple analytically solvable model, in section~\ref{sec:BTZ}, we consider the BTZ black brane which is asymptotically $AdS_3$. 
In the discussion section~\ref{sec:discussion} we provide a summary of our results and discuss promising extensions of our work. In particular, we consider possible implications of our results for quantum chaos and the bounds on quantum chaos~\cite{Maldacena:2015waa}. 
More details about the relevant calculations are provided in the appendices~\ref{N_H_expansion} and \ref{sec:Critical_points}. 

\section{Complexified quasinormal modes and pole-skipping}
\label{sec:dispersion_constraint}
	As our holographic model, we consider the Einstein-Hilbert action given by
	\begin{equation}
	S = - \frac{1}{16 \pi G_5} \int_{\mathcal{M}} d^5 x \,\,\sqrt{-g} \left(R - \frac{12}{L^2} \right)+ S_{bdy} \, ,
	\end{equation}
	with the $AdS$-radius, $L$. The Einstein equations in the trace-reduced form are given by
	\begin{equation}\label{EoM}
	R_{\mu \nu}+4 g_{\mu \nu}=\,0 \, .
	\end{equation}
	We use the usual scaling symmetries of the metric to set $L=1$. 
	Following the notation of~\cite{Policastro:2002se}, the solution dual to a thermal state is the Schwarzschild metric 
\begin{equation}\label{Metric_Gauge_u_coord}
ds^2=\frac{(\pi T)^2}{u}\left(-f(u)dt^2+\frac{}{}d \vec{x}^2\right)+\frac{1}{4u^2 f(u)}du^2 \, ,
\end{equation}
		with $ f(u)=1-u^2$ and  the Hawking temperature $T=r_h/\pi$, where $r_h$ is the location of the horizon in Poincar{\'e} coordinates. 

\subsection*{Scalar field in the bulk}
		In order to find the spectrum of long wavelength fluctuations of a boundary probe operator we study the quasi normal modes of the dual bulk field. We consider a scalar field of mass $m$ in the bulk.  The  corresponding equation of motion reads
		\begin{equation}\label{EoM_scalar}
		\bigg(	\partial_{\mu} (\sqrt{-g}\partial^{\mu} )- m^2 \sqrt{-g}\bigg)\Phi=\,0 \, .
		\end{equation}
		The field $\Phi(u)$ in the bulk is dual to a boundary operator $\mathcal{O}$ with the following conformal operator dimension~\cite{Witten:1998qj}\footnote{For example in a $d-$dimensioanl conformal theory, the conformal dimension of the charge density operator is $\Delta_J=d-1$ while that of the energy-momentum tensor operator is $\Delta_T=d$.}
		\begin{equation}\label{Delta}
		\Delta=2 \left(1+\sqrt{1+\mathfrak{m}^2}\right) \, .
		\end{equation}
		With the Fourier (plane wave) ansatz $\Phi(u,x_{\mu})=\int d^4 k \,  e^{- i \omega t+ i k z} \, \phi(u; \omega, k)$, the equation of motion becomes
		\begin{equation}\label{EoM_scalar_field}
		\phi''+\frac{u f' -f }{u f}\, \phi'+\frac{u \mathfrak{w}^2-(\mathfrak{m}^2+ \mathfrak{q}^2 u)f}{ u^2 f^2}\,\phi=\,0 \, ,
		\end{equation}
		where $\mathfrak{w}=\omega/2\pi T$, $\mathfrak{q}=k/2 \pi T$ and $\mathfrak{m}=\frac{m}{2}$.
		To find the spectrum of QNMs, we look for a bulk solution which is ingoing at the horizon. It can be expressed in terms of a near horizon expansion of the form
		\begin{equation}
		\phi(u)=(1-u)^{-i \mathfrak{w}/2}\sum_{n=0}^{\infty}a_n(\mathfrak{w}, \mathfrak{q}; \mathfrak{m}) (1-u)^n \, .
		\end{equation}
		Then the Dirichlet condition at $u=0$ gives the spectral curve of the operator
		\begin{equation}\label{sum}
		\sum_{n=0}^n a_n(\mathfrak{w}, \mathfrak{q}; \mathfrak{m}) =\,0:=\,\mathcal{S}(\mathfrak{q}^2, \mathfrak{w}) \, .
		\end{equation}
		Numerically	solving the spectral curve gives the spectrum of the QNMs associated with the operator of the scaling dimension $\Delta$ \eqref{Delta} \cite{Nunez:2003eq}.
The result associated with $\Delta=6$ is shown in figure~\ref{fig:dispersion}.

Before closing this section, let us  comment on the sum in  eq.~\eqref{sum}.		 According to \cite{Nunez:2003eq}, we need a sufficiently large but finite number of terms in the sum, so that by any change in that number of terms, the frequencies of the QNMs remain stable   up to a  specific number of digits. For the purposes  of this work, we have found that $n=50$ is sufficient to find the QNMs at the four lowest levels, in complete agreement with those of \cite{Nunez:2003eq}.

\subsection*{Pole-skipping}
In order to find the pole-skipping points of the operator $\mathcal{O}$ dual to the scalar field in the bulk, it is convenient to work with ingoing Eddington-Finkelstein coordinates. We first take $u=(\pi T/r)^2$ and then consider
\begin{equation}
dv= dt+\frac{dr}{r^2 f(r)}\, .
\end{equation}
Then the metric \eqref{Metric_Gauge_u_coord} can be rewritten as  
	\begin{equation}\label{Metric_Eddin]gton}
	ds^2=-r^2 f(r)dv^2+2dv dr +r^2 d\vec{x}^2 \, .
	\end{equation}
		The strategy is to find a near-horizon solution to the equation \eqref{EoM_scalar}. 
		Let us choose the Fourier ansatz $\Phi (r)= \int d^4 k \, e^{-i \omega v+ i k z} \, \phi(r;\omega,k)$  with $\phi(r;\omega,k)$ expanded near the horizon ($r_h=\pi T$) as 		
		\begin{equation}\label{phi_expanded}
		\phi(r)=\sum_{n=0}^{\infty}\phi_{n}(r-r_h)^n=\,\phi_0+(r-r_h)\phi_1+\cdots \, .
		\end{equation}
		Then by plugging \eqref{phi_expanded} back into \eqref{EoM_scalar},  we find a set of coupled algebraic equations for the set of coefficients $\phi_n$.  For instance, the first four of these equations can be formally written as 
			
		\begin{eqnarray}
		0&=&M_{11}\phi_0+4(\pi T)\,(i \wn-1)\phi_1\, ,\\
		0&=&M_{21}\phi_0+M_{22}\phi_1+8(\pi T)^2\,(i \wn-2)\phi_2\, ,\\
		0&=&M_{31}\phi_0+M_{32}\phi_1+M_{33}\phi_2+12(\pi T)^3\,(i \wn-3)\phi_3\, ,\\
		0&=&M_{41}\phi_0+M_{42}\phi_1+M_{43}\phi_2+M_{44}\phi_3+\,16(\pi T)^4\,(i \wn-4)\phi_4 \, ,
		\end{eqnarray}
		where the coefficients, $M_{rs}$, are in fact functions of $\wn$  and $\qn^2$.
		The resulting equations are special in the sense that at  $\ell^{th}$ order  of the near-horizonn expansion, one finds a linear equation relating $\phi_0$, $\phi_1$, $\cdots$ and $\phi_{\ell+1}$, with the coefficient of $\phi_{\ell+1}$ vanishing at the pole-skipping frequency $\wn=-i  (\ell+1)$. Here $T$ is the Hawking temperature of the black brane solution.
		
		As obvious from the above equations, just at the frequency $\wn_{\ell}=-i  \ell$, the first $\ell$ equations decouple from the rest of them and take the following form
		\begin{equation}
		0	=\,\mathcal{M}_{\ell\times \ell}(\wn=-i \ell,\qn^2_*)\begin{pmatrix}
		\phi_0\\
		\phi_1\\
		.\\
		.\\
		\phi_{\ell-1}\\
		\end{pmatrix}.
		\end{equation} 
		The roots of the equation $\det \mathcal{M}_{\ell\times \ell}(\wn=-i\ell, \qn^2_*)=0$,  are those wavenumbers at which,  for a given UV normalization constant, the ingoing boundary condition at the horizon is not sufficient to uniquely fix  a solution for $\Phi$ in the bulk. Let us call the roots $\qn_1^*, \qn_2^*, \cdots, \qn_{2\ell}^*$. Let us also recall that $\Phi$ is dual to the boundary operator $\mathcal{O}$.  It is clear that at these $2\ell$ points, the response function of the boundary operator, namely $G^R_{\mathcal{O}\mathcal{O}}(x_1-x_2)$, is multi-valued. These points are the so-called \textbf{\textit{pole-skipping points}}~\cite{Blake:2017ris,Blake:2018leo}.
		
		One concludes that to every pole-skipping frequency $\wn_{\ell}=-i  \ell$, $2\ell$ pole-skipping points of the dual boundary operator correspond:
		\begin{equation}\label{pole_skipping}
	\wn=\,-i \ell:\,\,\,\,\,\qn*=\{\pm\qn^*_{1,\ell}, \pm\qn^*_{2,\ell}, \cdots, \pm\qn^*_{\ell,\ell}\}
		\end{equation}
For later requirements, it is useful to express the pole-skipping points in terms of $\Delta$, the conformal dimension of the boundary operator $\mathcal{O}$. For the pole-skipping points at $\ell=1,2$, by using \eqref{Delta}, we find:
\begin{equation}
\begin{split}
\wn=-i:\,\,\,\,\,\,\,\qn^*_{1,1}=&\frac{1}{2}\,i\,\sqrt{\Delta^2-4\Delta+6}\, ,\\
\wn=-2i:\,\,\,\,\,\,\,\qn^*_{j,2}=&\frac{1}{2}\,i\,\sqrt{\Delta^2-4\Delta+12+\,(-1)^j2\sqrt{2}\sqrt{\Delta^2-4\Delta+6}},\,\,\,\,j=1,2 \, .
\end{split}
\end{equation}
The pole-skipping points at higher frequencies can be similarly computed through the near horizon analysis (see appendix \ref{N_H_expansion}), however, their analytic expressions are lengthy and we do not explicitly write them down here.
In the rest of this section we will  discuss the pole-skipping points and specifically how their behavior changes with $\Delta$. 
\begin{figure}
	\centering
	\includegraphics[width=0.46\textwidth]{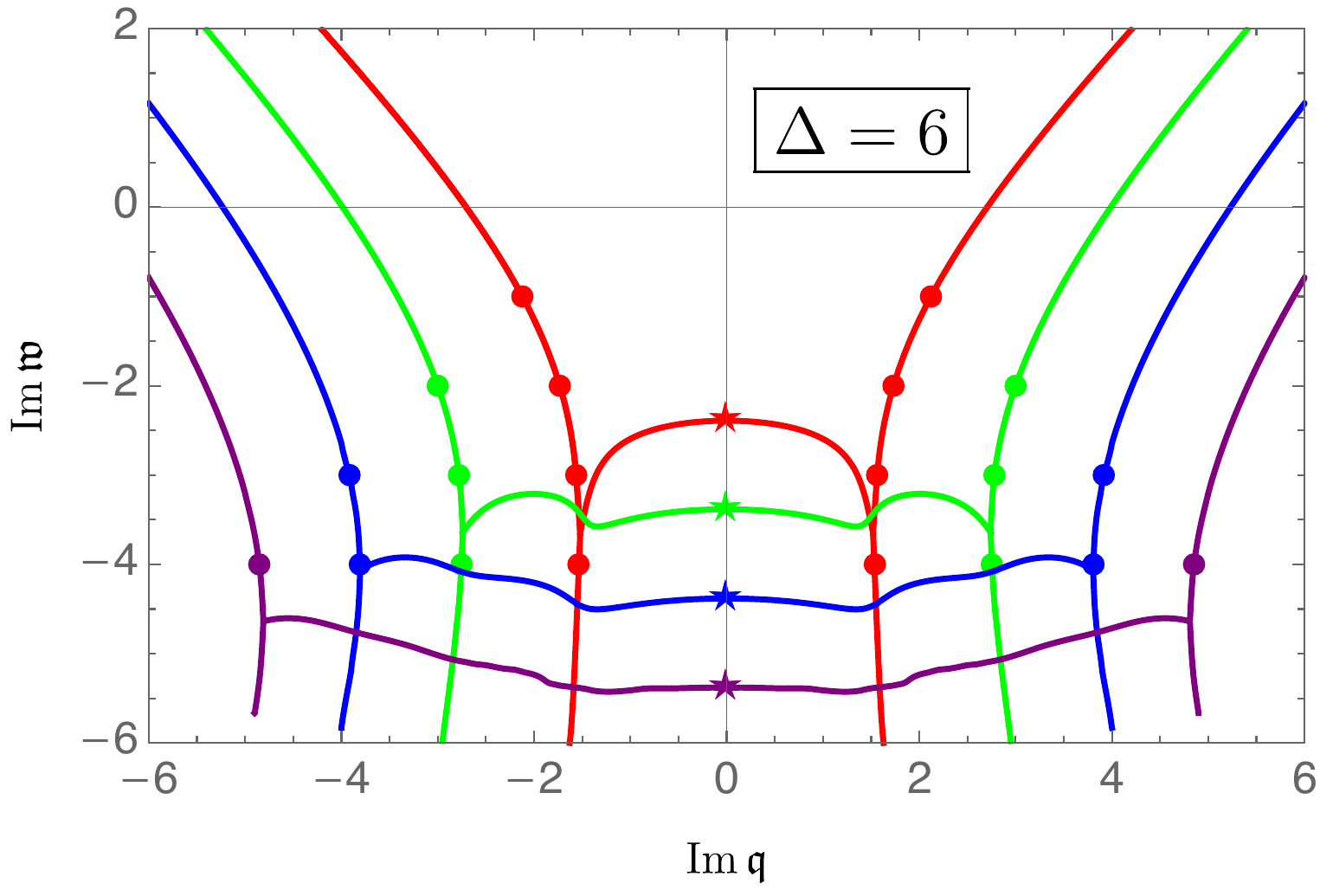}
	\includegraphics[width=0.46\textwidth]{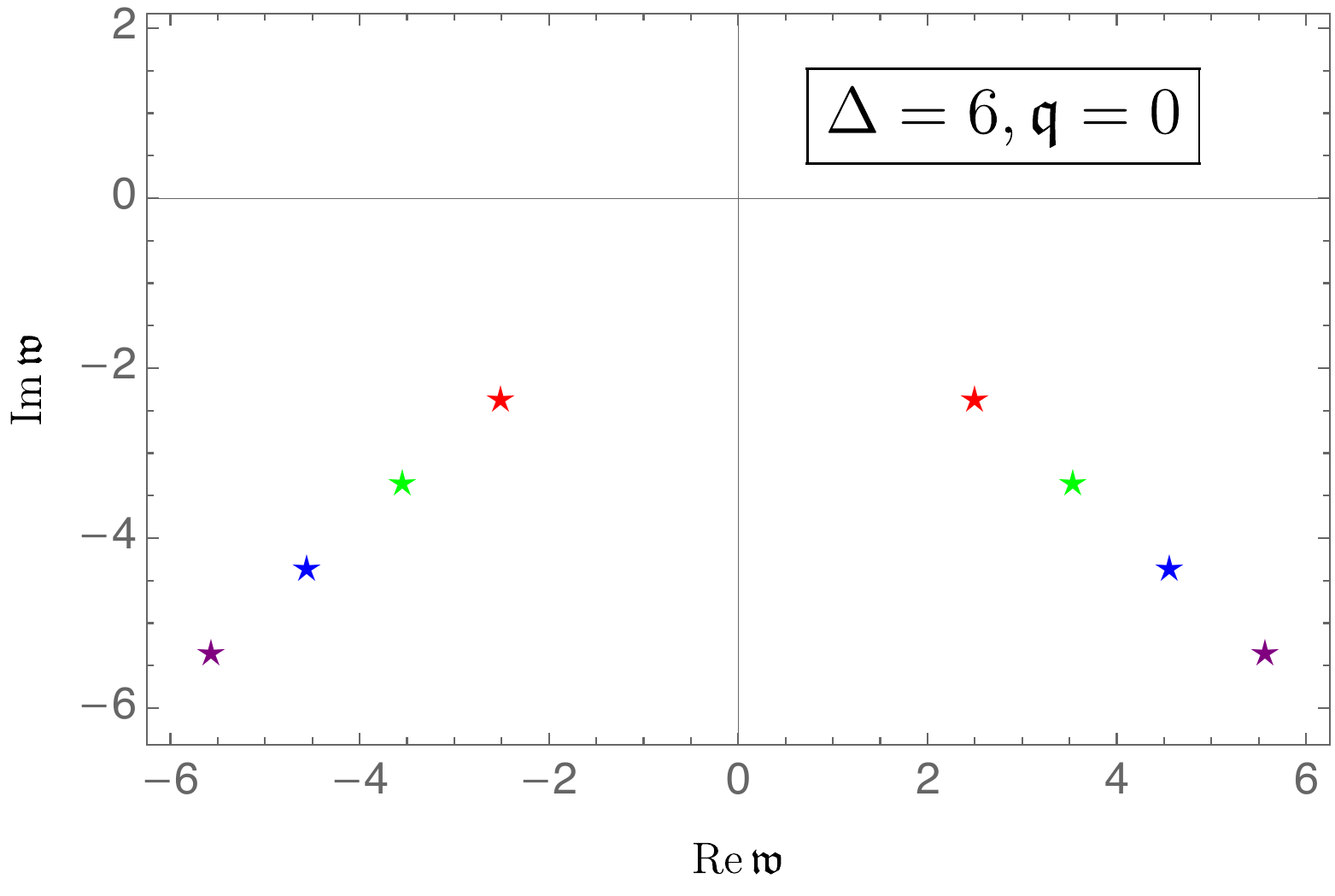}
	\caption{Left panel: Massive scalar dispersion relations for the lowest four QNMs evaluated at purely imaginary momenta, i.e.~$\text{Re}\mathfrak{q}=0$. The stars lying on the vertical axis denote the imaginary part of the QNMs at $\qn=0$. Colorful dots show the pole-skipping points at  frequencies $\text{Im}\,\wn=-1, -2, -3, -4$. Right panel: spectrum of the lowest four QNMs at $\qn=0$. These poles are exactly the poles indicated by stars in the left panel. 
	We have checked that this same figure arises for all values of $\Delta$ as well as for massive vector and tensor fields which will be discussed in section~\ref{sec:vectorTensor}.}
	\label{fig:dispersion}
\end{figure}

Let us begin by pointing out a very interesting property of pole-skipping points which is common for all values of $\Delta$. To that end, we go back and revisit figure~\ref{fig:dispersion}.
The colorful dots in the left plot are the pole-skipping points  associated with the response function of a boundary operator with $\Delta=6$. It is observed that all pole-skipping points lie on the dispersion relations of QNMs at imaginary momenta.
 Although this result has not been well known for a theory without gapless excitations, it is in agreement with previous studies~\cite{Blake:2019otz,Grozdanov:2019uhi}. 
 
However, the non-trivial feature of our result is revealed when splitting the pole-skipping points lying on  different dispersion relations.
This behavior has not been shown for gapped modes before. 
As it is obvious in the figure, we indicate which pole skipping points belong to which QNM dispersion relation by showing them in the same color. Each color indicates a different QNM dispersion relation and the corresponding set of pole-skipping points:
\begin{equation}\label{Red_Green_Blue_Purple}
\begin{split}
\text{Red dots:}&\,\,\,(\pm\qn^*_{1,1}, -i), (\pm\qn^*_{1,2}, -2i), (\pm\qn^*_{1,3}, -3i), (\pm\qn^*_{1,4}, -4i), \cdots \, ,\\
\text{Green dots:}&\,\,\,\,\,\,\,\,\,\,\,\,\,\,\,\,\,\,\,\,\,\,\,\,\,\,\,\,\,\,\,\,\,(\pm\qn^*_{2,2}, -2i), (\pm\qn^*_{2,3}, -3i), (\pm\qn^*_{2,4}, -4i), \cdots\, ,\\
\text{Blue dots:}&\,\,\,\,\,\,\,\,\,\,\,\,\,\,\,\,\,\,\,\,\,\,\,\,\,\,\,\,\,\,\,\,\,\,\,\,\,\,\,\,\,\,\,\,\,\,\,\,\,\,\,\,\,\,\,\,\,\,\,\,\,\,\,\,\,\,\, (\pm\qn^*_{3,3}, -3i), (\pm\qn^*_{3,4}, -4i),\cdots\, ,\\
\text{Purple dots:}&\,\,\,\,\,\,\,\,\,\,\,\,\,\,\,\,\,\,\,\,\,\,\,\,\,\,\,\,\,\,\,\,\,\,\,\,\,\,\,\,\,\,\,\,\,\,\,\,\,\,\,\,\,\,\,\,\,\,\,\,\,\,\,\,\,\,\,\,\,\,\,\,\,\,\,\,\,\,\,\,\,\,\,\,\,\,\,\,\,\,\,\,\,\,\,\,\,\,\,\,  (\pm\qn^*_{4,4}, -4i),  \cdots \, .
\end{split}
\end{equation}
The location of the red dots can be described by a ``one-to-one" map between the red dots and discrete frequencies  $\omega=-m\,i(2\pi T)$ with the {\it pole-skipping levels} labeled by $m=1,2, \cdots$. This means that the lowest QNM with {\it mode number} labeled by $n=1$, namely the one with the dispersion relation shown in red, is constrained to coincide with the pole-skipping frequencies for all levels $m\ge 1$ with $m\in \mathbb{N}$. But the dispersion relation of the second level QNM meets one less constraint; the Green dispersion relation is constrained by the pole-skipping points with $m\ge2$. By every step going one level higher in the tower of QNMs, the number of pole-skipping frequencies constraining the dispersion relations decreases by one.

Let us emphasize that figure 1 does not show the $\text{Re}\,\wn$. That just shows $\text{Im}\,\wn$ at \textit{purely imaginary momenta}: $\qn^2=e^{i \theta}|\qn^2|=\,-|\qn^2|$. 
In other words all points in the figure 1 correspond to $\theta=\pi$. In order to show why the classification of curves by the colors  used in the figure 1 is \textit{correct}, we illustrate  the position of poles (by poles we mean the stars shown in the right panel) in the complex $\wn$ plane at $\theta=\pi$.

\begin{figure}
	\centering
	\includegraphics[width=0.5\textwidth]{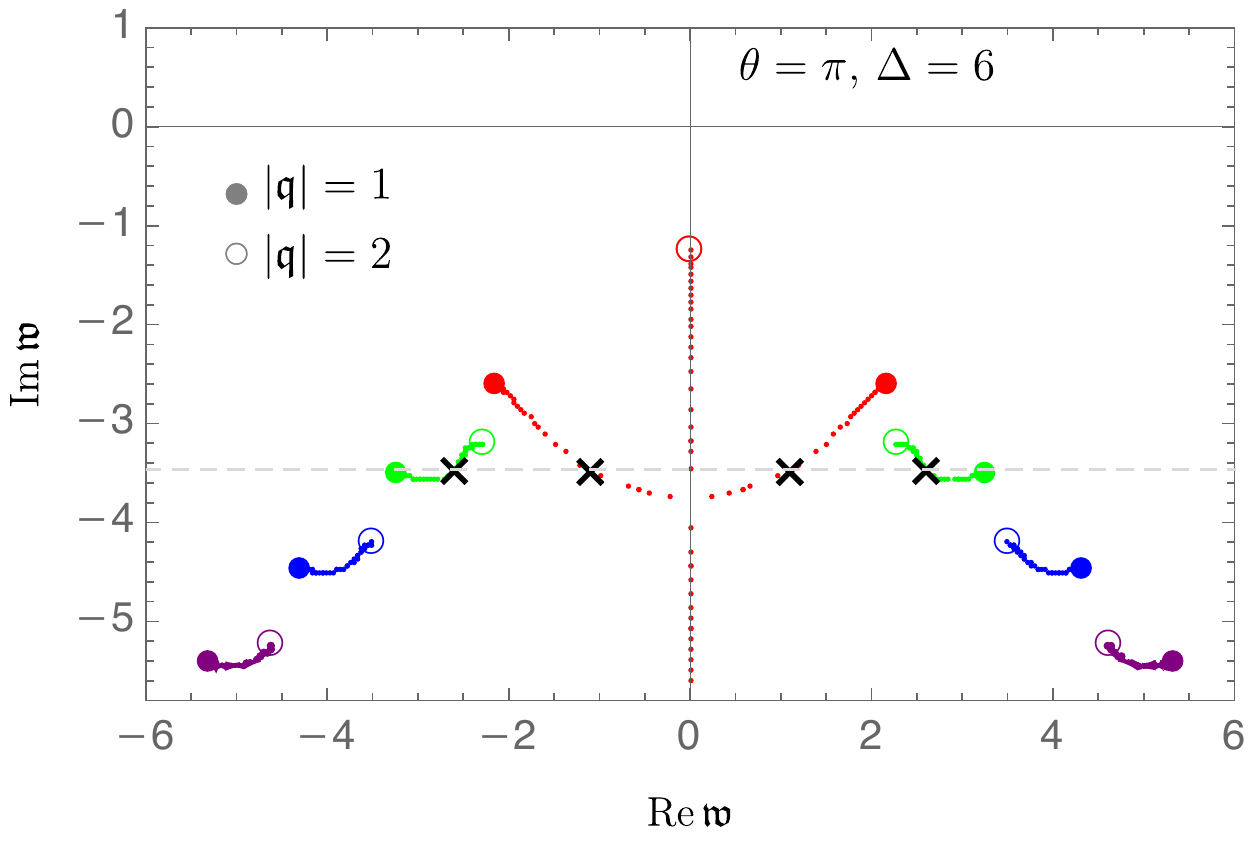} 	
	\caption{This figure shows the same QNMs as figure 1, however, now in the complex frequency plane.  
		Along the trajectories, $|\mathfrak{q}^2|$ changes from 1 (dots) to 2 (circles).
		This range is chosen in order to resolve the apparent collision between the red and green pole in figure 1 and the apparent branch point of the red curve in that same figure, both occuring near $\text{Im}\,\wn\approx -3.5$. All modes are shown for the fixed phase $\theta=\pi$. }
	\label{fig:JHEP}
\end{figure}

The result is given in figure~\ref{fig:JHEP}. Each pole is identified with a dot at $|\qn|=1$. Now we start to vary $|\qn|$ in the interval $(1,2)$. When the imaginary $\qn$ increases, namely when $|\qn|$ increases, 
with fixed $\theta=\pi\sout{=\text{constant}}$, each pole moves along a trajectory and arrives at a circle. \textit{It is obvious from the plot that in the range $|\qn|\in(1,2)$ the modes $n=1$ and $n=2$   do not collide at $\theta=\pi$.} 
The horizontal dashed line in figure~\ref{fig:JHEP} shows that at $\text{Im}\,\qn=1.50$ (the middle point of the interval), the green and red stars have the same $\text{Im}\,\wn=-3.49$ (at the locations indicated by the black crosses). This is exactly what is observed in figure~\ref{fig:dispersion}. But in figure~\ref{fig:JHEP} we see that the real part, $\text{Re}\,\wn$, for the red and green stars is not the same.  
This simply reveals that the crossing between the red and green curves in figure 1 is not a mode collision point. 

More importantly, the above discussion makes it clear how we have chosen the colors in the figure~\ref{fig:dispersion}. 
They are selected such that one color, e.g. Red, indicates the trajectory of one QNM at a particular level, e.g. the red one is the lowest-lying QNM.

This counting also works for the constraints on hydrodynamic modes in  a theory with macroscopic conserved currents. By definition, hydrodynamic modes are constrained to pass through $(\qn=0, \wn=0)$. For a non-hydrodynamic QNM, however, $\wn(\qn\rightarrow 0)$ is a complex frequency with negative imaginary part. 
In this sense, the hydrodynamic modes are restricted by all $m\ge 0$, as shown for the diffusion and shear mode in~\cite{Blake:2019otz}. 
According to this analysis, non-hydrodynamic QNMs are not as severely constrained as the hydrodynamic modes at real $\qn$. On the other hand, according to the absolute value of $\wn(\qn\rightarrow 0)$, they are classified into different levels in a tower spectrum, including the lowest-lying level, the first excited level and so on.
Then the behavior observed in figure~\ref{fig:dispersion} suggests that there is a relation between this classification and the constraints on their dispersion relations at imaginary $\qn$'s: \newline 
\textit{The more pole-skipping points lie on the dispersion relation of a gapped QNM, the more constrained that mode is at imaginary $\qn$'s. Then our observation implies that the longest-lived mode in the system is the one which is the most constrained. The latter is actually the lowest-lying QNM. To summarize, our results indicate 
that the hierarchy in the life-time of QNMs originates from the number of constraints induced by the pole-skipping points at imaginary momenta.}

At this point, we emphasize that the results obtained from figure~\ref{fig:dispersion}  all consider purely imaginary momenta. It should also be noted that the dispersion relations for QNMs at imaginary or complex-valued momentum do not directly tell us anything about the dispersion relations of the same QNMs at real-valued momentum. Hence we can not derive a rigorous statement for the relation between the hierarchy of QNM life-times (at real-valued momenta) and the number of pole-skipping points (at imaginary/complex-valued momentum). Hence we stress that our statement above is an observation.

\section{Derivative expansion for gapped modes}
\label{sec:gapped}
There is yet another interesting feature associated with the dispersion relations shown in figure~\ref{fig:dispersion} which deserves attention. 
\textbf{\textit{Critical points}} have been defined as those points at which two QNMs ``collide'' with each other~\cite{Grozdanov:2019kge,Grozdanov:2019uhi}. This means that the real and imaginary parts of their frequencies agree with each other at a particular value of a complex momentum. This is illustrated in the upper panel of figure~\ref{fig:Scalar_cross_before}, \ref{fig:Scalar_cross_beyond} and~\ref{fig:Scalar_cross_at}, which we will discuss further below. Looking at the dispersion relation of each QNMs shown in figure~\ref{fig:dispersion}, one type of potential critical points can immediately be recognized; the points where the dispersion relation of the QNM splits into two branches. Such points correspond to the collision of two modes with the same $\text{Im}\,\qn$. However, they might be some other critical points that cannot be specified in Im $\wn$ - Im $\qn$ plane.

It should be noted that 
at a critical point, $\text{Re}\, \mathfrak{w}$, $\text{Re}\, \mathfrak{q}$, $\text{Im}\mathfrak{w}$ and $\text{Im}\mathfrak{q}$ of the two colliding QNMs have to be identical. We discuss this point further below.

Very recently, in the context of hydrodynamics, it has been shown that the \textit{\textbf{critical points of the  spectral curves}} defining a given hydrodynamic QNM may determine the convergence radius of the derivative expansion of that QNM. There, the spectral curve is defined to be an implicit function relating the frequency and momentum, i.e.~defining the dispersion relation of the mode. 
This definition of critical points utilizing the spectral curve yields the same values as the definition through the collision of QNMs mentioned above~\cite{Grozdanov:2019kge}. In other words, QNMs collide at critical points of the spectral curve.

The scalar field fluctuation we choose to consider now has no hydrodynamic mode. 
In other words, the corresponding spectral curve does not include any dispersion relation passing through the origin. 
Instead, in this section we intend to study the analytic structure of the spectral curve with the aim of establishing a relation between critical points and the 
gapped QNMs displayed in figure~\ref{fig:dispersion}. 
Eventually, this will lead us to the convergence radii of derivative expansions for gapped modes. 

The spectral curve of a given operator $\mathcal{O}$ can be written in Fourier space as
\begin{equation}\label{eq:spectral}
\mathcal{S}(\mathfrak{w}, \mathfrak{q}^2)=\,0
\end{equation}
where both $\mathfrak{w}$ and $\mathfrak{q}^2$ are complex numbers. 
The solution to this (implicit function) equation is a discrete set  of QNMs $\wn^{(n)}\equiv\wn^{(n)}(\qn^2)$. In our present case:
\begin{equation}
\lim_{\qn^2\rightarrow 0}\wn^{(n)}(\qn^2)=\wn^{(n)}_{\text{g}} \, ,
\end{equation}
where $\wn^{(n)}_{\text{g}}$ is complex with $\text{Im}\,\mathfrak{w}^{(n)}_\text{g}<0$ corresponding to the (generally complex-valued) gap 
of the $n^{th}$ mode. Among the QNMs, the one with smallest $|\text{Im}\,\wn_{g}|$ is the longest lived mode in the system. Obtaining any information about this mode may reveal how fast the operator is thermalized. 
In the previous section, we have shown results in figures~\ref{fig:dispersion} and \ref{fig:JHEP} which display information obtained numerically by solving eq.~\eqref{eq:spectral} through the standard method of computing QNMs by requiring ingoing boundary conditions at the horizon and Dirichlet boundary condition at the conformal boundary. 
In this section, we want to extract information from the spectral curve by studying its analyticity properties. To this end, we search perturbative solutions to~\eqref{eq:spectral} in the long wavelength limit.  
Formally, one writes the solution in a derivative expansion in momentum space\footnote{Let us emphasize that this is not a hydrodynamic derivative expansion; equation \eqref{sol} does not pass through the origin. This is a derivative expansion about the $n^{th}$ gapped QNM. In this sense, it differs from what has been discussed in \cite{Withers:2018srf} and \cite{Grozdanov:2019kge}.}
\begin{equation}\label{sol}
\mathfrak{w}^{(n)}= \mathfrak{w}_{\text{g}}^{(n)}-i\,\sum_{n=1}^{\infty} a_k^{(n)} \mathfrak{q}^{2k} \, .
\end{equation}
The set of coefficients $a_k^{(n)}$ corresponds to the branch of the {\it Puiseux series}~\cite{Grozdanov:2019kge,Grozdanov:2019uhi} passing through  the point $( \mathfrak{w}_{\text{g}}^{(n)}, 0)$ \footnote{In holography, we do not know the structure of $\mathcal{S}(\mathfrak{w}, \mathfrak{q}^2)$ in general. However in well-known examples, it does not feature any non-analyticity, such as that of RTA kinetic theory~\cite{Heller:2020hnq}. Thus, according to the implicit function theorem, one may expect the behavior of $\mathcal{S}(\mathfrak{w}, \mathfrak{q}^2)$ to be described by a Puiseux series with a non-zero radius of convergence.}.
 In general, one can perturbatively find the coefficients $a_{k}^{(n)}$ for each of the gapped modes in the system. But as we will explain in detail, we limit our study to the two cases $n=1,\,2$. 

\begin{figure}
	\centering
	\includegraphics[width=0.75\textwidth]{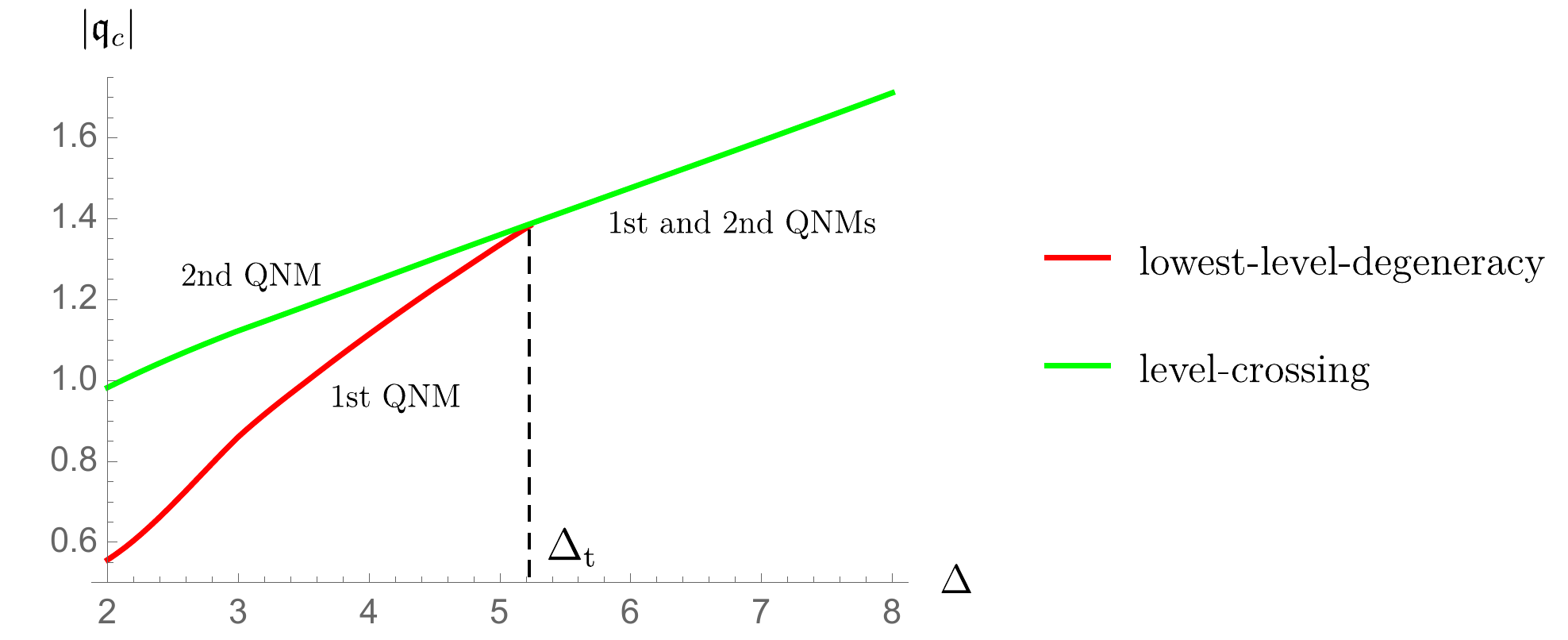} 	
	\caption{The radius of convergence, $|\mathfrak{q}_c|$, of the derivative expansion versus the scaling dimension of the boundary operator, within the range $2 \le \Delta \le 8$. The transition between the regime of level-crossing and the regime of lowest-level-degeneracy occurs at $\Delta_\text{t}\approx 5.23$.
	}
	\label{fig:two_phase}
\end{figure}

More precisely, we would like to  investigate to what extent the derivative expansions  associated with $\wn^{(1)}$ and $\wn^{(2)}$ are convergent. Thus we follow  \cite{Grozdanov:2019kge} and  compute the corresponding critical  points of the spectral curve. 
For this purpose one needs to find the set of roots 
$\{\text{Re}\, \mathfrak{w},\, \text{Im}\, \mathfrak{w},\, \text{Re}\, \mathfrak{q}^2, \, \text{Im}\, \mathfrak{q}^2\}$ satisfying the following two complex-valued equations\footnote{It is worth mentioning that, in general, the radius of convergence about $\mathfrak{w}_{\text{g}}^{(n)}$ is determined by the singularity closest to it~\cite{Heller:2020hnq}. 
In this sense,  eqs.~\eqref{eq_set} may admit solutions which do not necessarily correspond to singularities of  $\wn^{(n)}(\qn)$ \cite{Heller:2020hnq} (see ~\sec{BTZ_section} as an example illustrating this point). However, in our present case, solving these equations gives the correct radius of convergence.}
\begin{equation}\label{eq_set}
\mathcal{S}(\mathfrak{q}^2, \mathfrak{w})=\,0,\,\,\,\,\,\,\,\,\,\,\,\,
\frac{\partial \mathcal{S}(\mathfrak{q}^2, \mathfrak{w})}{\partial \mathfrak{w}}=\,0.
\end{equation}
Starting with $\mathfrak{w}_{\text{g}}^{(1)}$, we look for the   critical point  obtained from \eqref{eq_set} which has the smallest distance to  $(0, \mathfrak{w}_{\text{g}}^{(1)})$.  Let us call this critical point $( \qn_c^{(1)}, \wn_c^{(1)})$. Then $|\qn_c^{(1)}|$  sets the radius of convergence of the derivative expansion about $\mathfrak{w}_{\text{g}}^{(1)}$. We can repeat the computation by starting with  $\mathfrak{w}_{\text{g}}^{(2)}$ in order to find $|\qn_c^{(2)}|$, the radius of convergence of the derivative expansion about the second quasinormal  mode, $\mathfrak{w}_{\text{g}}^{(2)}$. 

We have found $|\qn_c^{(1)}|$ and $|\qn_c^{(2)}|$ for operators of weight $2\le\Delta\le8$. The result is shown in figure~\ref{fig:two_phase}.
As it can be seen in the figure, it turns out that $|\qn_c^{(2)}|$ is approximately a linear function of $\Delta$. In contrast to that, the curve showing $|\qn_{c}^{(1)}|$ as a function of $\Delta$ has a kink at a particular operator dimension $\Delta=\Delta_\text{t}$:
%
\begin{equation}\label{eq:Critical_Delta}
\Delta_{\text{t}}\,\approx\, 5.23 \, .
\end{equation}
When $\Delta<\Delta_{\text{t}}$, we find $|\qn_{c}^{(1)}|<|\qn_{c}^{(2)}|$, while for $\Delta>\Delta_{\text{t}}$, the two derivative expansions about  the  lowest QNM ($n=1$) or the first excited QNM ($n=2$) have the same radius of convergence: $|\qn_{c}^{(1)}|=|\qn_{c}^{(2)}|$.

\begin{SCfigure}[\sidecaptionrelwidth]
	\centering
	\includegraphics[width=0.5\textwidth]{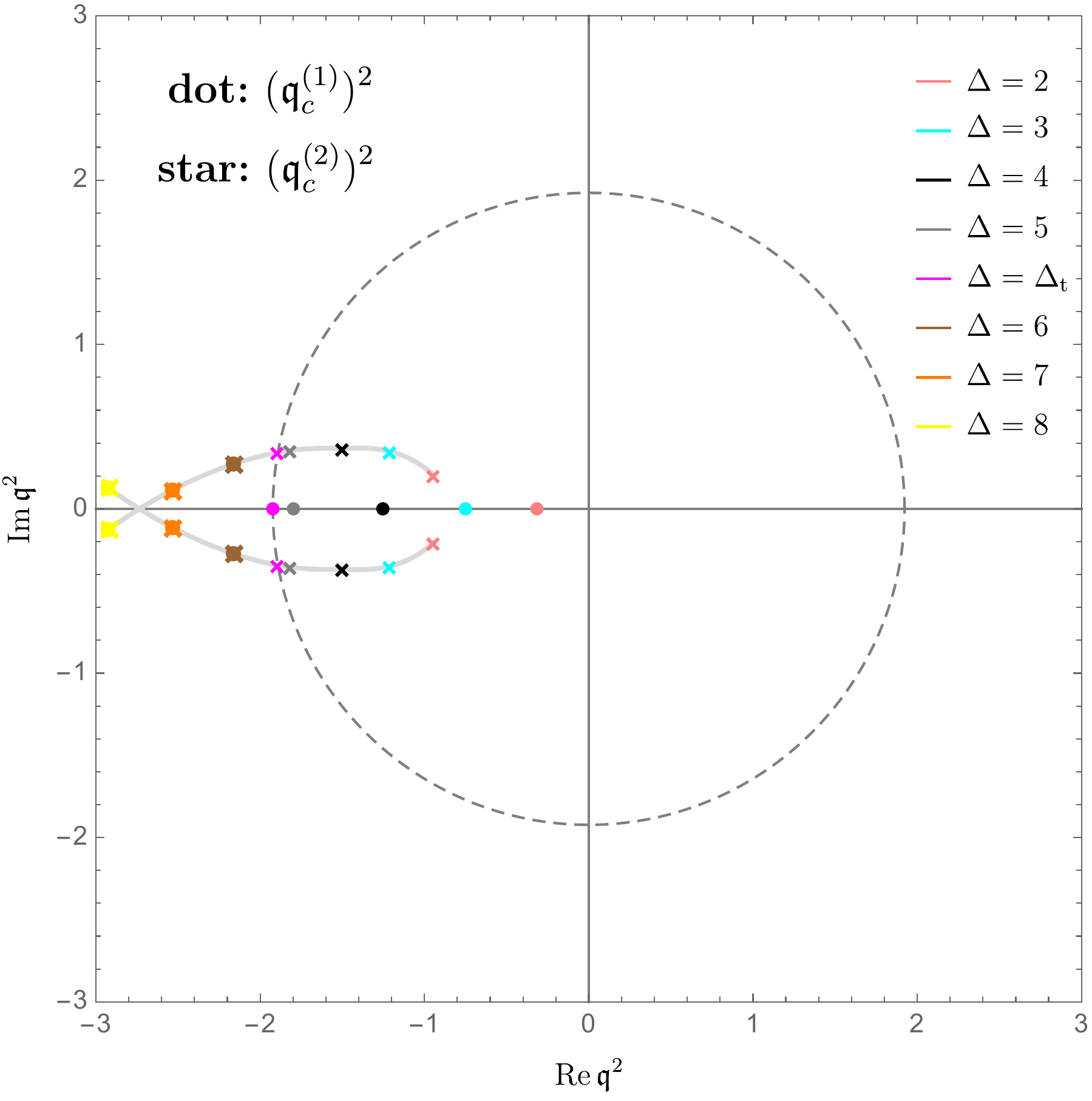} 	
	\caption{The location of $\qn^2_c$ in the complex plane for various values of $\Delta$ within the range $2\le\Delta\le 8$. Every dot (cross), corresponding to a particular value of $\Delta$, lies on a circle whose  center is at the origin; the region enclosed by the circle is then the domain of convergence of the  derivative expansion about the  first (second) QNM associated with that value of $\Delta$. The dashed circle shows a transition between two regimes: for values of $\Delta<\Delta_{\text{t}}$, the dot does not lie on the circle passing through the crosses. In addition,  dots are located in the $(\text{Im}\,\qn^2)$-axis.  But for $\Delta>\Delta_{\text{t}}$, dots and crosses lie on the same circle.\,\,\,\,\,\,\,\,\,\,\,\,\,\,\,\,\,\,\,\,\,\,\,\,\,\,\,\,\,\,\,\,\,\,\,\,\,\,\,\,\,\,\,\,\,\,\,\,\,\,\,\,\,\,\,\,\,\,\,\,\,\,\,\,\,\,\,\,\,\,\,\,\,\,\,\,\,\,\,\,\,\,\,\,\,\,\,\,\,\,\,\,\,\,\,\,\,\,\,\,}
	\label{fig:circle}
\end{SCfigure}

\begin{figure}[tb]
	\centering
	\includegraphics[width=0.45\textwidth]{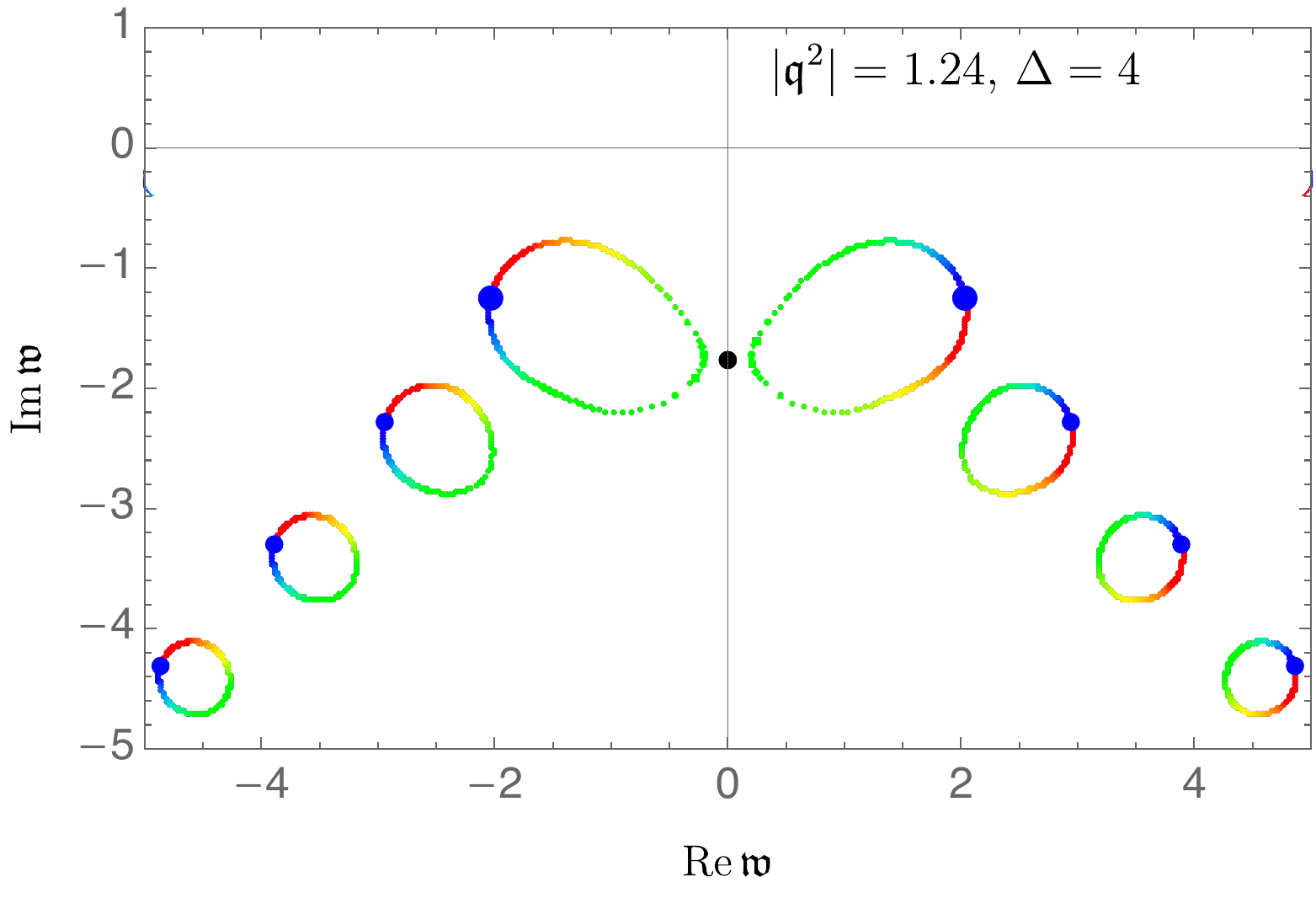}\,\,\,\,\,\,\,\,\includegraphics[width=0.45\textwidth]{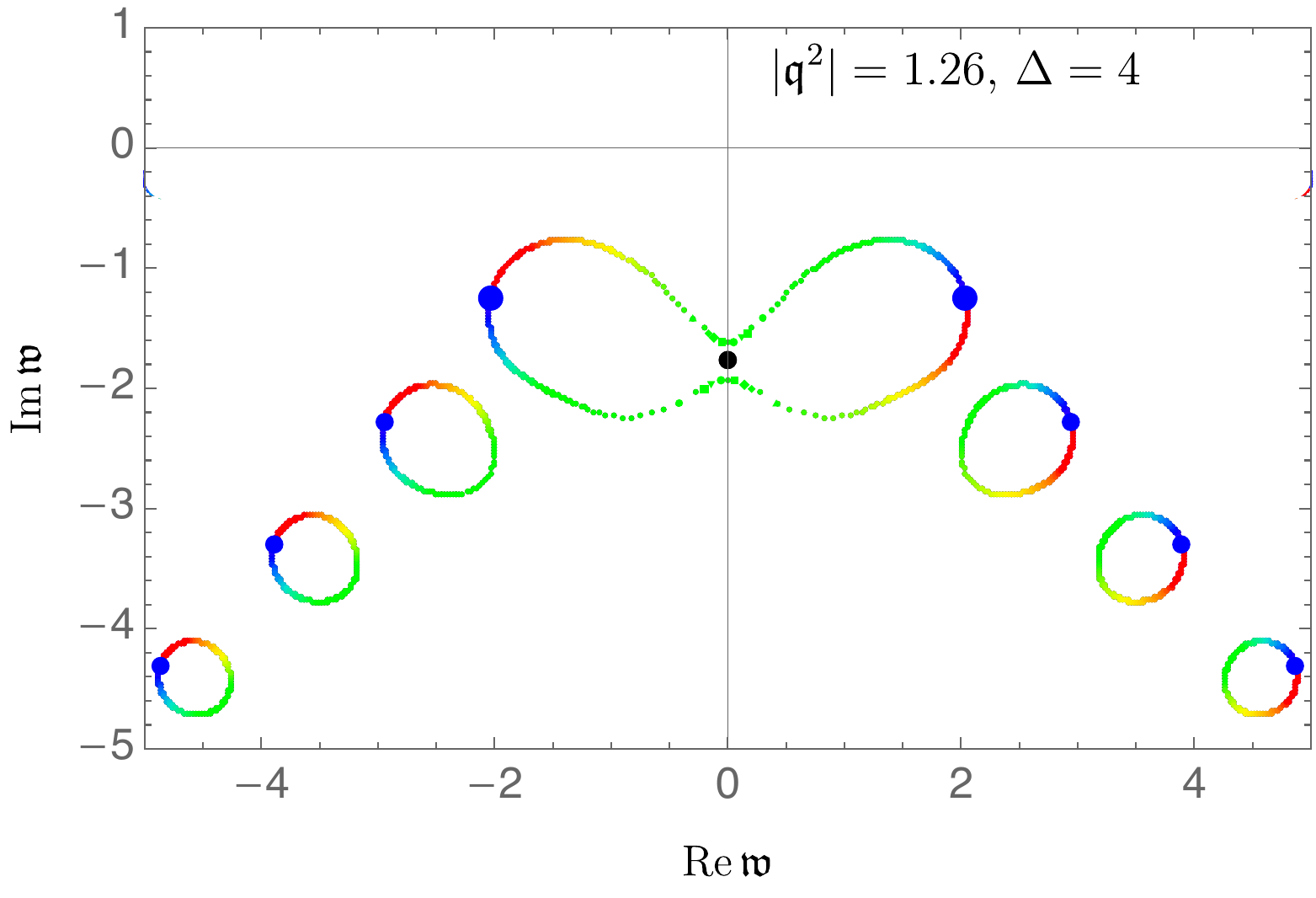} 	
	\includegraphics[width=0.45\textwidth]{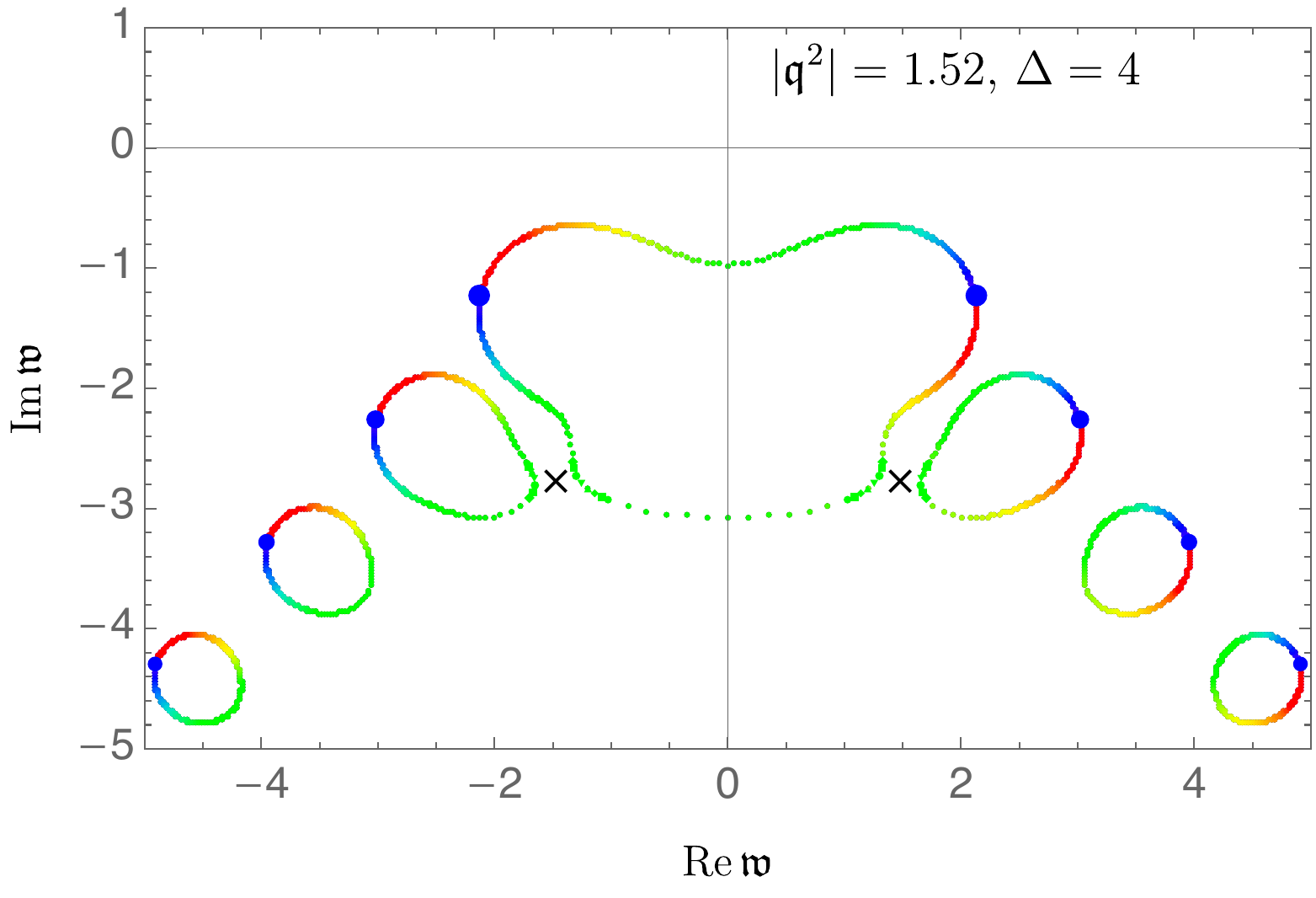}\,\,\,\,\,\,\,\,\includegraphics[width=0.45\textwidth]{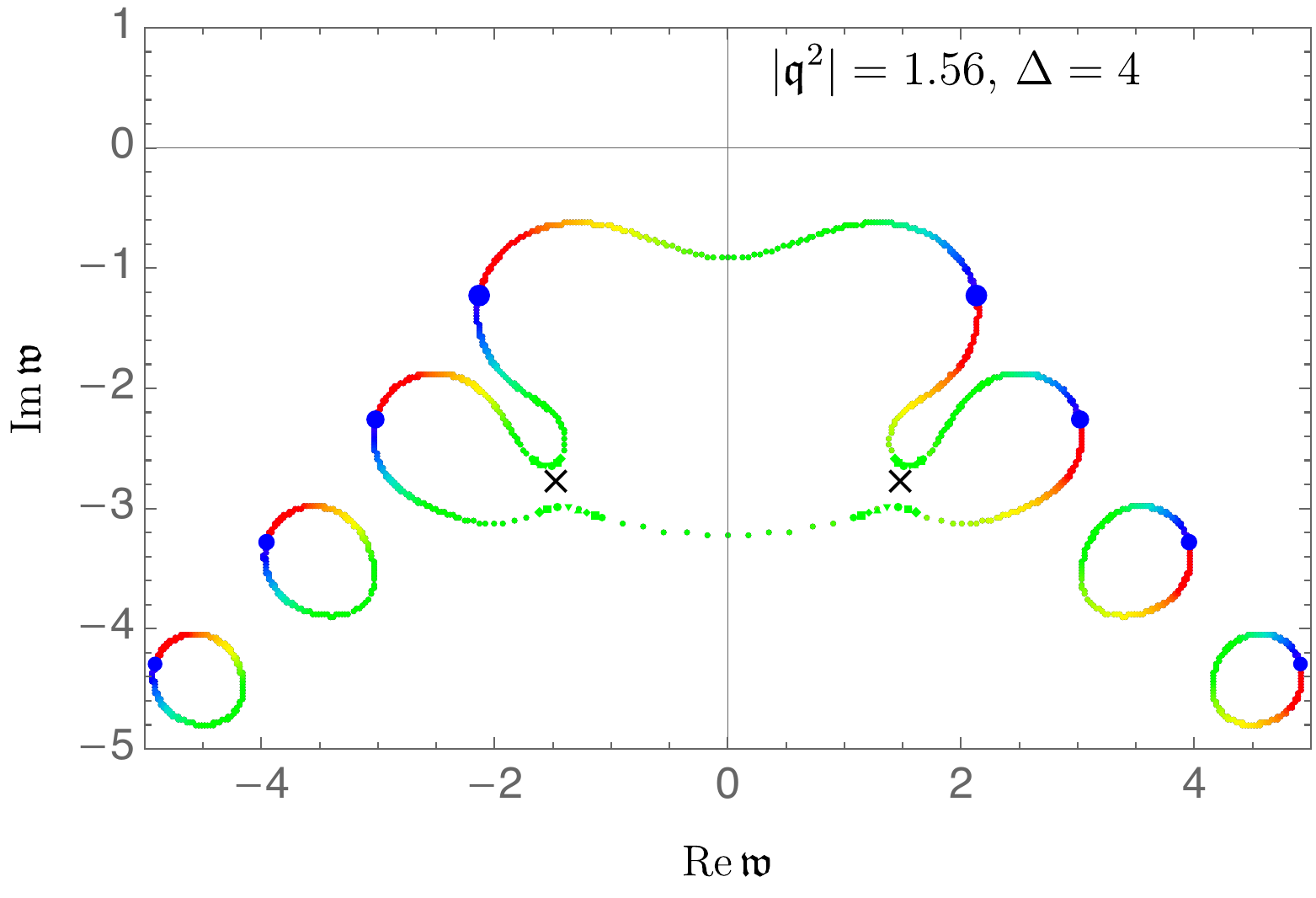} 	
	\caption{
	\textit{Lowest-level-degeneracy.} Poles of the retarded two-point function of the scalar operator, in the complex $\wn-$plane, at various values of
		the complexified momentum $\qn^2=|\qn^2|e^{i \theta}$.   Blue dots correspond
		to the location of  poles for real $\qn^2$, namely for $\theta=0$.  As $\theta$ increases from $0$ to $2\pi$, each pole moves counter-clockwise	following the trajectory whose color changes continuously from blue to red. The two top panels show the situation slightly before and after the collision point marked by black dots. At the critical value $|\qn_c^2|\approx 1.25$, the trajectories of the lowest-lying modes collide. After the collision, the orbits of these modes  are no longer closed: the two of them exchange their positions as the phase $\theta$ increases from $0$ to $2\pi$. The second collision occurs at a higher momentum.  The two bottom panels show the situation slightly before and after the second collision, marked by black crosses. At the critical value $|\qn_c^2|\approx 1.54$, the trajectories of the second QNMs 
collide with the common trajectory of the lowest-lying modes. After the collision, the two lowest-lying modes together with the second QNMs 
exchange their positions cyclically as the phase $\theta$ increase from $0$ to $2\pi$. }
	\label{fig:Scalar_cross_before}
\end{figure}

\begin{figure}[tb]
	\centering
	\includegraphics[width=0.45\textwidth]{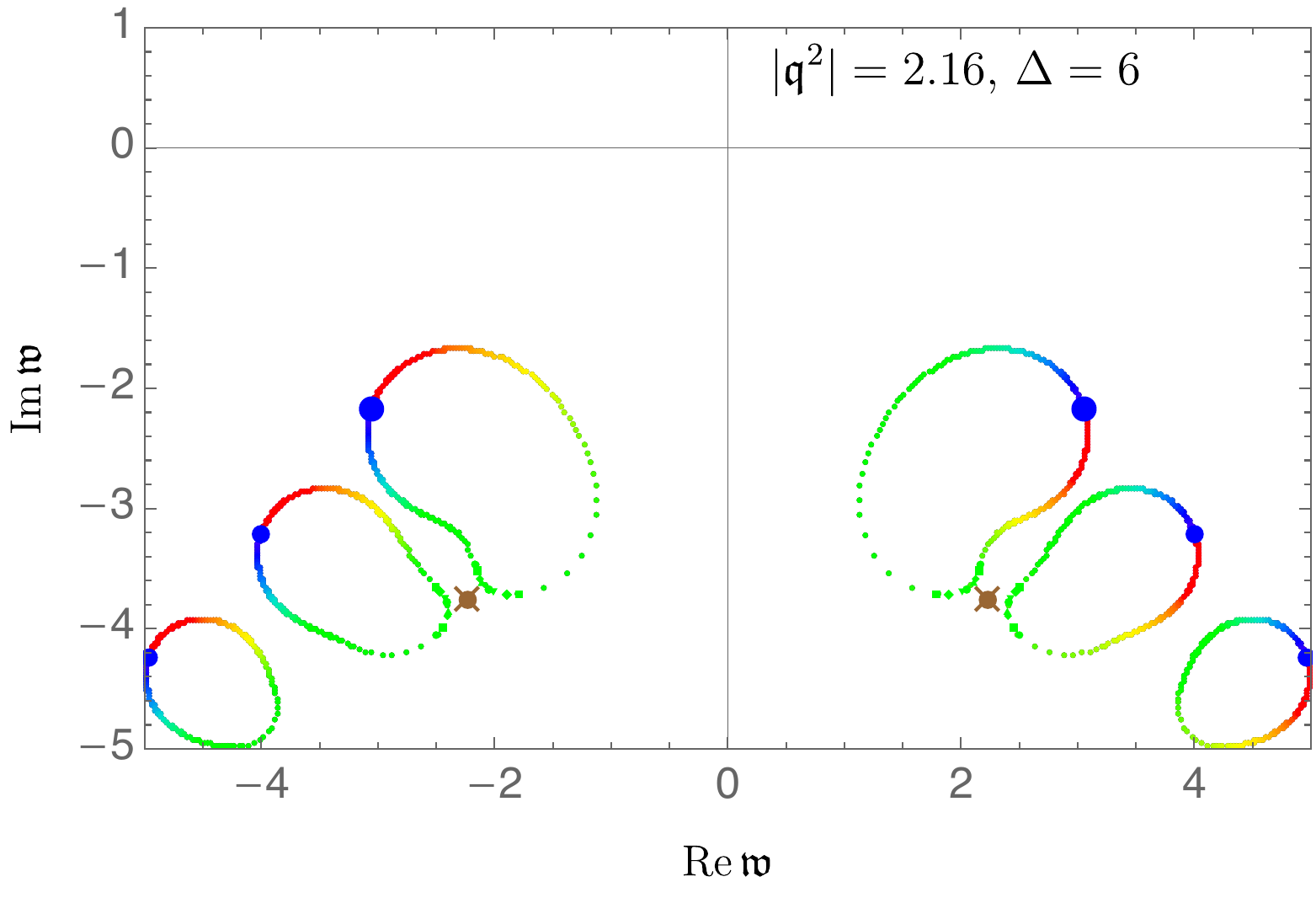}\,\,\,\,\,\,\,\,\includegraphics[width=0.45\textwidth]{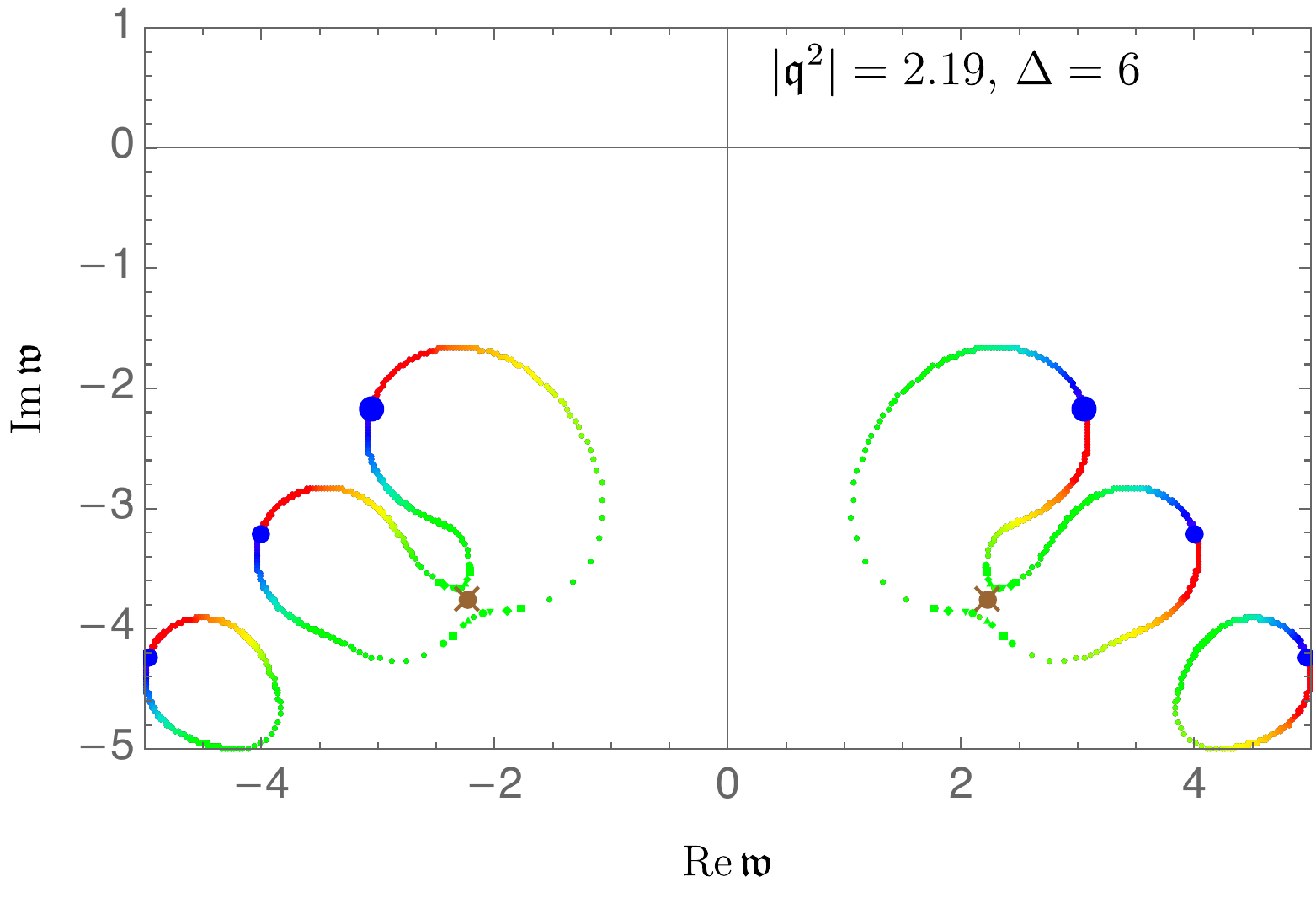} 	
	\caption{
	\textit{Level-crossing. } 
	Poles of the retarded two-point function of the scalar operator, in the complex $\wn-$plane, at various values of
		the complexified momentum $\qn^2=|\qn^2|e^{i \theta}$. Blue dots correspond
		to the location of  poles for real $\qn^2$, namely for $\theta=0$. As $\theta$ increases from $0$ to $2\pi$, each pole moves counter-clockwise following the trajectory whose color changes continuously from blue to red. The two  panels show the situation slightly before and after the collision point marked by brown dot-crosses. Before the collision the orbits of all QNMs are closed. At the critical value $|\qn_c^2|\approx2.18$, the trajectories of the lowest-lying modes collide with those of the second QNMs.  
		After the collision,  orbits of mentioned modes  are no longer closed: in both left and right sides of the plane, the lowest-lying and the second QNMs 
exchange their positions as the phase $\theta$ increases from $0$ to $2\pi$.}
	\label{fig:Scalar_cross_beyond}
\end{figure}

\begin{figure}[tb]
	\centering
	\includegraphics[width=0.45\textwidth]{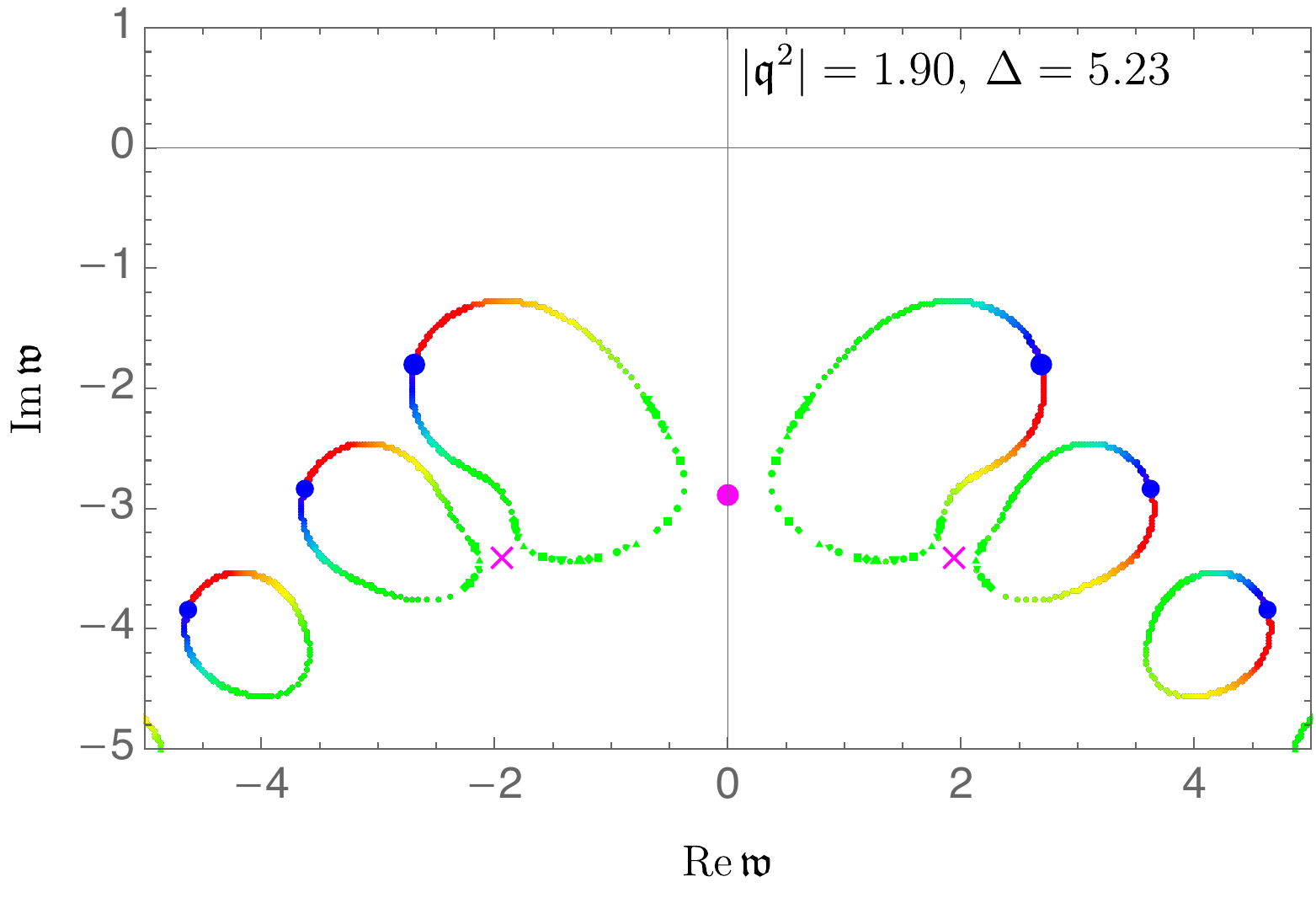}\,\,\,\,\,\,\,\,\includegraphics[width=0.45\textwidth]{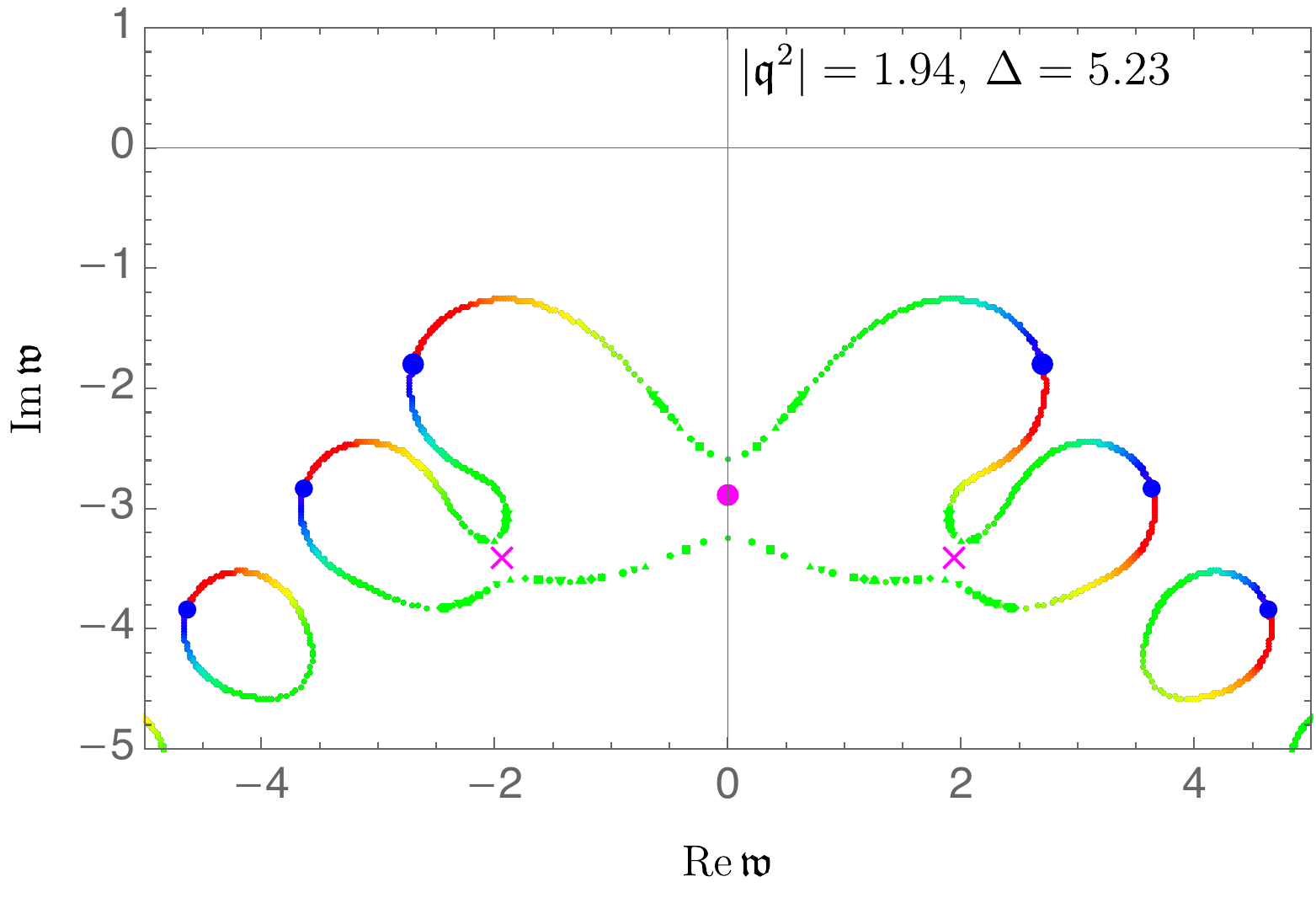} 	
	\caption{
	\textit{Simultaneous level-crossing and lowest-level-degeneracy at transition. } 
	This figure shows the same modes as figures~\ref{fig:Scalar_cross_before} and \ref{fig:Scalar_cross_beyond} but now at the intermediate value of the conformal operator dimension $\Delta= 5.23$. That is approximately the transition dimension $\Delta_\text{t} $ which separates the two types of collision regimes, level-crossing versus lowest-level-degeneracy, from each other. It is apparent that now both types of collisions appear simultaneously. Before the collision, in the left panel, the two lowest-lying QNMs ($n=1$) have two separate trajectories, and the two second QNMs ($n=2$) have two separate trajectories. These four trajectories then collide with each other simultaneously at the three critical points indicated by the magenta dot and the magenta crosses. These collision points are also displayed in magenta in the complex momentum plane in figure~\ref{fig:circle}.
	}
	\label{fig:Scalar_cross_at}
\end{figure}

The transition between  the two regimes discussed above can also be visualized in a different way. Since in our model $\wn\equiv\wn(\qn^2)$,  we can display the critical points in the complex $\qn^2$-plane. We have done so for various values of $\Delta$ in figure~\ref{fig:circle}. For every $\Delta<\Delta_{\text{t}}$, the convergence of the derivative expansion about the lowest QNM is identified with only one collision point which lies on the horizontal axis, i.e.~at $\text{Im}\,\mathfrak{q}^2=0$. Such points are indicated by dots in the figure and occur at purely imaginary momenta. The distance between each dot and the origin sets the radius of convergence. Within the same range of $\Delta$,  to each value of $\Delta$ two crosses correspond, too. The pair of crosses are located symmetrically with respect to the $(\text{Im}\,\qn^2)$-axis and set the domain of convergence of the derivative expansion about the 
second QNMs. 
The latter is actually set by the radius of the circle centered at the origin passing through the two crosses.
%
%
At $\Delta=\Delta_{\text{t}}$, the dot and the two crosses are all located at the same distance to the origin. One is to conclude that at this special value of $\Delta$, the radii of convergence about the lowest-lying and the second QNM are the same but are determined by distinct points in the complex $\qn^2$-plane. For $\Delta>\Delta_{\text{t}}$, however, in each side of the $(\text{Im}\,\qn^2)$-axis, there is a dot and  a cross coinciding with each other. It simply shows that beyond the transition value of $\Delta$, the radius of convergence of the derivative expansion about the lowest-lying and the second QNMs is set by the same value and is identified with the same critical point. See appendix~\ref{sec:Critical_points} for a detailed discussion, and figures~\ref{fig:Collision_Delta_4} and~\ref{fig:Collision_Delta_6} for illustrations of the collision points on the complex-valued dispersion relations.
%
\par\bigskip 
\noindent

The existence of a transition value for $\Delta$ has an important outcome for the complex life of the lowest-lying QNMs.  Indeed, it 
distinguishes two distinct types of scattering between complexified QNMs depending on whether $\Delta<\Delta_{\text{t}}$ or $\Delta>\Delta_{\text{t}}$. Each type of scattering point determines the critical point $( \qn_c^{(1)}, \wn_c^{(1)})$ within the respective regime of $\Delta$. 
To clarify this point, we demonstrate the collision points for two values of $\Delta$ in figures~\ref{fig:Scalar_cross_before}, \ref{fig:Scalar_cross_at} and \ref {fig:Scalar_cross_beyond}.
\par\bigskip 
\noindent
%

Let us first consider figure~\ref{fig:Scalar_cross_before} which corresponds to $\Delta=4<\Delta_{\text{t}}$. As was discussed earlier, in this range, radius of convergence for the  lowest-lying modes differs from that of the second QNMs. 
For this reason, in the figure,  we have separated the two cases. In the two top panels we show the positions of complexified modes slightly before (left plot) and after (right plot) the first collision. 
The collision occurs at $|\qn^2_c|\approx1.25$ setting the radius of convergence about the lowest-lying QNM to be $|\qn_c^{(1)}|\approx(1.25)^{1/2}$. 
\footnote{This value agrees with the result given in the tensor sector of metric fluctuations given in~\cite{Grozdanov:2019uhi}, associated with the energy-momentum tensor which has $\Delta=4$. } This collision point is actually the black dot already introduced in figure~\ref{fig:circle} and as one expects,  corresponds to the point with $\Delta=4$ on the left red branch of figure~\ref{fig:two_phase} as well. Since at the collision point the two lowest-lying modes coincide with each other, we refer to it as the \textbf{\textit{lowest-level-degeneracy}}.

By increasing $|\qn^2|$, the second QNMs (with mode number $n=2$) collides with the first QNM.
In the two bottom panels of figure~\ref{fig:circle}, we have shown the situation slightly before and after such a collision. 
The collision occurs at $|\qn^2_c|\approx1.54$ at points marked by two black crosses in the figure. So  $|\qn_c^{(2)}|\approx(1.54)^{1/2}$ sets the radius of convergence for the derivative expansions about  the second QNMs. 
The collision points are actually the black crosses already introduced in figure~\ref{fig:circle} and as one expects,  correspond to the point with $\Delta=4$ on the  green branch of figure~\ref{fig:two_phase}. 

As an example for $\Delta>\Delta_{\text{t}}$, we show the case $\Delta=6$ in figure~\ref{fig:Scalar_cross_beyond}. As was discussed earlier, in this range, the radius of convergence for the  lowest-lying modes is equal to that of the second QNMs. So one expects one single type of collision between the trajectories of complexified modes to solely determine the radius of convergence.  In figure~\ref{fig:Scalar_cross_beyond},  we indicate the positions of complexified modes slightly before and after such a collision, marked by brown dot-crosses.
The collision occurs at $|\qn^2_c|\approx2.18$ setting the radius of convergence about all four of the lowest QNMs to be $|\qn_c^{(1)}|=|\qn_c^{(2)}|\approx(2.18)^{1/2}$.  
The two collision points  are actually the brown dot-crosses already introduced in figure~\ref{fig:circle} and as one expects,  correspond to the point with $\Delta=6$ on the plot of figure~\ref{fig:two_phase}. Since at each collision point one of the lowest-lying QNMs collides with the respective second QNM,  we refer to this collision as \textbf{\textit{level-crossing}} \cite{Grozdanov:2019kge}.

\par\bigskip 
\noindent
Finally, at the transition point $\Delta_\text{t}$, shown in figure~\ref{fig:Scalar_cross_at},at $|\qn_c^2|$, three collisions occur. 
So a level-crossing occurs simultaneously with a lowest-level-degeneracy. 
By ``simultaneously occur'', we mean that they correspond to the same $|\qn_c^2|$, although not at a common $\theta$. 
From the field theory point of view, this constellation of poles may be viewed as the defining feature, fixing the value of $\Delta_\text{t}=5.23$. An analytic understanding and physical interpretation of this value remains to be understood. This same constellation of poles at $\Delta_\text{t}$ is illustrated in the complex momentum plane in figure~\ref{fig:circle}. The magenta dot and crosses there indicate the same poles as the magenta dot and crosses in figure~\ref{fig:Scalar_cross_at}. 

In summary, we find that the transition first observed in figure~\ref{fig:two_phase} is translated to a transition from the lowest-level-degeneracy to the level-crossing behavior of the QNMs at complex momenta. As shown by the magenta dots and crosses in figure~\ref{fig:circle}, just at the transition value of $\Delta$, the lowest-level-degeneracy and the level-crossing occur simultaneously. Thus they jointly determine the radius of convergence about the lowest-lying QNM.

\section{Bound on the radius of convergence from pole-skipping }
\label{bound}
In section~\ref{sec:dispersion_constraint} we saw how the pole-skipping points constrain the dispersion relation of QNMs at imaginary momenta. In this section we will show that the pole-skipping points also constrain the radius of convergence of the derivative expansion about  the lowest-lying QNMs.

\begin{figure}
	\centering
	\includegraphics[width=0.9\textwidth]{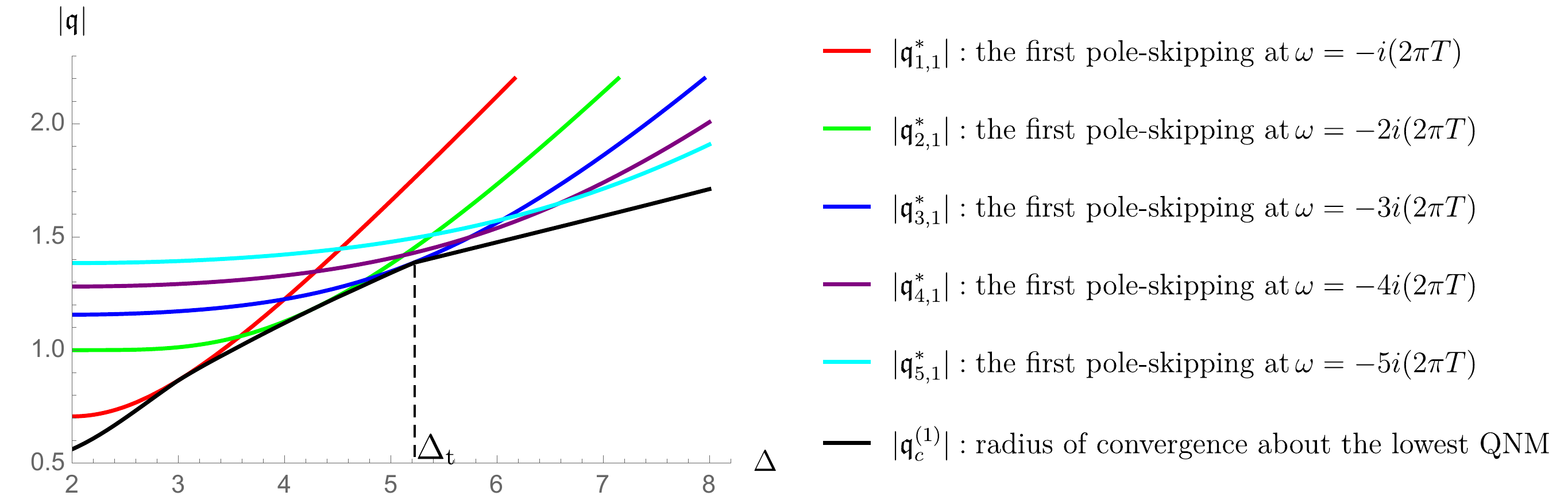} 
	\caption{Comparison between the radius of convergence of the derivative expansion about the lowest-lying gapped QNM, determined by $|\qn_c^{(1)}|$, and the smallest absolute value of the momentum at pole-skipping points. The red, green and blue curves can be continued to larger values of $\Delta$, but here we focus on the lower values of $|\qn|$.
		Higher QNMs with $n=4,\, 5$ have been considered and do not intersect or coincide with the black line of $|\qn_c^{(1)}|$-values in the range $2\le\Delta\le 8$.
	}
	\label{fig:Scalar_cross}
\end{figure}

In figure~\ref{fig:Scalar_cross} we have compared the above mentioned convergence radius, namely $|\qn_c^{(1)}|$,  with the absolute value of the momenta associated with the pole-skipping points at frequencies: $\wn=-i, -2i, -3i$. More precisely, at each of these frequencies, we consider the pole-skipping point with the smallest absolute value of its momentum. In the language of \eqref{pole_skipping}, the red, green, blue, purple and cyan curves in figure~\ref{fig:Scalar_cross} correspond to $|\qn^*_{1,1}|$, $|\qn^*_{2,1}|$, $|\qn^*_{3,1}|$, $|\qn^*_{4,1}|$ and $|\qn^*_{5,1}|$, respectively. 

We observe that the black curve in figure~\ref{fig:Scalar_cross}, indicating the radius of convergence, $|\qn_c^{(1)}|$, is actually the envelope of the pole-skipping curves. In other words, 
the radius of convergence of the derivative expansion about the lowest-lying mode is  bounded from above. The upper-bound is given by the distance between the origin and that pole-skipping point which is closest to the origin. One may write this as 
\begin{equation}\label{eq:upper_bound}
|\qn_c^{(1)}|\le\text{min}\{|\qn^*_{1,1}|, |\qn^*_{2,1}|, |\qn^*_{3,1}|, \cdots\} \, .
\end{equation}
At specific values of $\Delta$ this bound is saturated. We find three such saturation points, namely those at which the red, green and blue curves coincide with the black curve in figure~\ref{fig:Scalar_cross}. One of the saturation values, interestingly, is the transition value $\Delta_{\text{t}}$ that we already discussed in   section~\ref{sec:gapped}. Recall that the black curve has a kink at exactly this transition value $\Delta_\text{t}$. One might guess that the transition from lowest-level-degeneracy to the level-crossing behavior of QNMs  is the consequence of the fact that  $|\qn_c^{(1)}|$ cannot exceed the bound \eqref{eq:upper_bound}. One may further infer that the kink in the black curve is also related to this bound. Analytic solutions would help to reveal these potential relations.

\section{Vector and tensor operators}
\label{sec:vectorTensor}
Our discussion so far has been limited to a scalar gapped operator. Similar arguments can be made for gapped operators of higher spins. In this section we do so in the case of vector and tensor operators in the boundary theory. 

Let us start with studying a vector operator $\mathcal{V}$; this operator is dual to a massive vector field $A^{\mu}$ in the bulk. The corresponding equation of motion is
 \cite{Aharony:1999ti,Kabat:2012hp}
\begin{equation}
\nabla_{\mu}F^\mu_\nu-m^2 A_{\nu}=\,0 \, .
\end{equation}
We consider transverse perturbations of this field around the AdS$_5$ black brane solution \eqref{Metric_Gauge_u_coord}:
\begin{equation}
A= A_\mu \, dx^\mu = \int d^4 k \, A_{x}(u;\omega, k)\, e^{- i \omega t +i k z} \, dx \, .
\end{equation}
Defining $E_x= -i\wn A_x$, we arrive at
\begin{equation}\label{E_x_eq}
E_x''+\frac{f' }{ f}\, E_x'+\frac{4u( \wn^2- \qn^2 f)- m^2 f}{ 4u^2 f^2}\,E_x=\,0 \, .
\end{equation}
Then the spectrum of QNMs associated with the field $E_x$ in the bulk corresponds to  non-hydrodynamic gapped modes of the operator $\mathcal{V}$, excited by $E_x$ on the boundary. The conformal weight of $\mathcal{V}$ is given in terms of $m$, namely the mass of $A^{\mu}$ in AdS$_{d+1}$, as follows
\begin{equation}\label{}
\Delta=\frac{d}{2}+\sqrt{\frac{(d-2)^2}{4}+m^2}\,\,\,\,\,\,\xrightarrow[]{d=4}\,\,\,\,\,\Delta=2+\sqrt{1+m^2} \, .
\end{equation}
Let us recall the unitarity bound~\cite{Minwalla:1997ka} on operators with the spin $\ell$,
\begin{equation}
\label{eq:unitarityBound} 
\Delta \ge d+\ell-2 \, . 
\end{equation}
So vector operators ($\ell =1$) in $d=(3+1)$ dimensions are bounded by $\Delta \ge 4+1-2=3$. 

\begin{figure}
	\centering
	\includegraphics[width=0.8\textwidth]{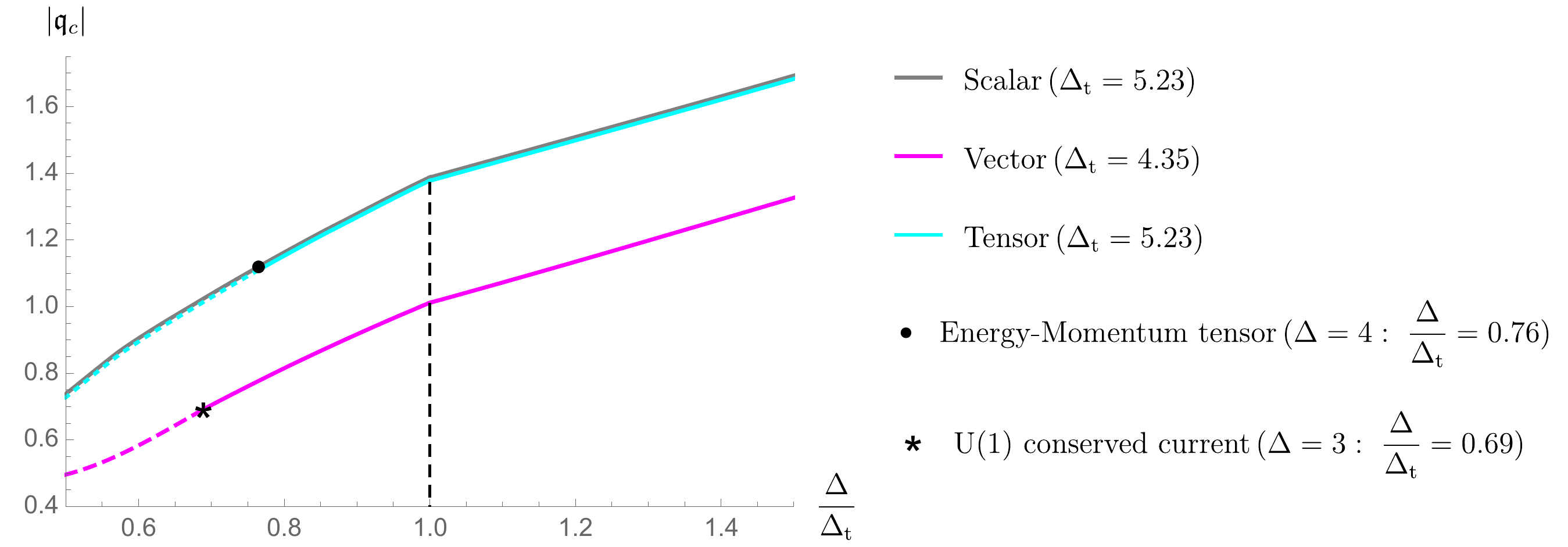} 	
	\caption{The radius of convergence of the derivative expansion about the lowest-lying QNMs versus the scaling dimension of the boundary operator, within the range $1/2 \le \Delta/\Delta_{\text{t}} \le 3/2$. The black dot, which corresponds to transverse perturbations of the metric in the bulk was found in \cite{Grozdanov:2019uhi}. 
		The dashed parts of the magenta and cyan curves show the values of $\Delta$ excluded by the unitarity bound for the vector and tensor, respectively.  
		The black curve for the scalar operator coincides over the whole range with the cyan curve for the tensor operator. However, for the scalar operator all values of $\Delta$ displayed here are allowed by the scalar's unitarity bound, $\Delta\ge 2$. 
	}
	\label{fig:delta_c}
\end{figure}

It turns out that the spectral curve of $E_x$ has similar properties to that of the scalar field in the bulk. We have checked that the dispersion relations of its QNMs  at imaginary momenta behave qualitatively just like those shown in figure~\ref{fig:dispersion}. Thus the statement that QNMs are constrained by pole-skipping points continues to hold for $\mathcal{V}$ in perfect analogy to the scalar operator case. The transition between the lowest-level-degeneracy and the level-crossing (see figure~\ref{fig:two_phase})  is observed in this case as well.   We find the transition value of $\Delta$ for vector operators to be 
\begin{equation}\label{eq:vectorTransitionDelta}
\Delta_{\text{t}}\approx 4.35 \, .
\end{equation}
In order to compare the mentioned transition with that of the scalar operator,  in figure~\ref{fig:delta_c}, we have considered operators of dimension $1/2< \Delta/\Delta_{\text{t}}< 3/2 $ with integer spins.
The black curve in figure~\ref{fig:Scalar_cross} is identical to the grey curve shown in figure~\ref{fig:delta_c}. 

For this vector operator we have confirmed that the pole-skipping points and the radius of convergence satisfy the same relations displayed for the scalar operator in figure~\ref{fig:Scalar_cross}. The figure for the vector is qualitatively the same and hence we do not show it here. However, the transition conformal dimension for the vector operator, given in eq.~\eqref{eq:vectorTransitionDelta}, differs from that of the scalar operator, given in eq.~\eqref{eq:Critical_Delta}.

In figure~\ref{fig:delta_c}, we have also shown the results concerning a gapped tensor operator, $\mathcal{T}$, in the boundary theory. This operator is dual to the transverse component of a   massive spin two field in the bulk. Let us refer to the massive spin two field as $\varphi_{\mu\nu}$. 
The corresponding equations of motion are given by~\cite{Benini:2010pr}:
\begin{equation}\label{}
\begin{split}
0=&\,(\Box-m^2)\varphi_{\mu\nu}+2 R_{\mu \lambda \nu \rho}\varphi^{\lambda \rho} \, ,\\
0=&\,D^{\mu}\varphi_{\mu\nu}\, , \\
0=&\,\varphi^{\mu}_{\mu} \, .
\end{split}
\end{equation}
The fluctuation of the $(xy)$-component decouples from the rest of the fluctuations. This $\varphi_{xy}$ sources the fluctuations of $\mathcal{T}$ on the boundary. In a $d$-dimensional boundary theory, the conformal dimension of the operator $\mathcal{T}$
	is given by 
	\begin{equation}
	\Delta=\frac{1}{2}(d+ \sqrt{d^2+4 m^2})\,\,\,\,\,\,\xrightarrow[]{d=4}\,\,\,\,\,\Delta=2+\sqrt{4+ m^2} \, .
	\end{equation}
Note that the unitarity bound~\eqref{eq:unitarityBound} forces the traceless tensor operator dimension to obey $\Delta\ge 4$.  
Taking
\begin{equation}\label{}
\varphi_{xy}=\int d^4 k\,\frac{\psi(u; \omega, k)}{u}\,e^{- i \omega t + i k z},
\end{equation}
it turns out that $\psi(u)\equiv \psi(u;\omega,k)$ obeys exactly the same equation that a massive scalar fields does obey in the bulk, namely \eqref{EoM_scalar_field}. 
Hence, the horizon expansion and location of pole-skipping points for the tensor sector are identical to those presented in the scalar sector above.

Thus, as expected, we find the QNMs of $\mathcal{T}$ to feature exactly the same behavior as the scalar operator's QNMs. This can be seen in figure~\ref{fig:delta_c}. The gray and cyan curves coincide with each other.  The black dot shown on the cyan curve corresponds to $\mathcal{T}$ with $\Delta=4$ (or equivalently with $\Delta/\Delta_{\text{t}}=0.76$). An example for such an operator is the shear component of the energy-momentum tensor, namely $T_{xy}$. 
For this $\Delta=4$ we find $|\qn_c|=1.11942$ which is in agreement with the critical point momentum computed from the spin-2 tensor perturbations of the metric, $h_{xy}$, provided in equation (4.38a) of~\cite{Grozdanov:2019uhi} as well as with figure 13 in that same paper.

Let us note another feature displayed in figure~\ref{fig:delta_c} which is related to the unitarity bound~\eqref{eq:unitarityBound}. For vector and tensor operators the minimum allowed value of $\Delta$ is identical to the $\Delta$ of the conserved vector and tensor operators, namely the $U(1)$ current with $\Delta=3$ and the energy-momentum tensor with $\Delta=4$, respectively. 
It is remarkable that the upper bound~\eqref{eq:upper_bound} is obeyed by the vector operator for all values in agreement with the unitarity bound $\Delta\ge 3$ which we have checked. However, if the unitarity bound is violated, also the bound~\eqref{eq:upper_bound} is violated.

\section{Comments on the BTZ case ($AdS_3$)}
\label{BTZ_section}
\label{sec:BTZ}
Our analysis can be simply extended to the $AdS_{d+1}$ case with $d>2$. In three bulk dimensions, however, the situation is different. In this case, one studies the QNMs around the BTZ black hole. 

From a non-extremal BTZ black hole, the retarded Green's function of  an operator with the conformal dimension $\Delta$ in the boundary CFT can be computed analytically. For  a non-integer $\Delta$ \cite{Son:2002sd}:
 %
  \begin{equation}\label{} 
  \begin{split}
G^{R}(\wn, \qn)=&\,C_{\Delta}\,\Gamma \left(\frac{\Delta}{2}+ \frac{i (\wn -\qn)}{2}\right)\,\Gamma \left(\frac{\Delta}{2}- \frac{i (\wn -\qn)}{2}\right)\,\Gamma \left(\frac{\Delta}{2}+ \frac{i (\wn +\qn)}{2}\right)\\
&\times  \Gamma \left(\frac{\Delta}{2}-\frac{i(\wn+\qn)}{2}\right)\bigg[\cosh (\pi \qn)- \cos(\pi \Delta)\cosh(\pi \wn)+ i \sin(\pi \Delta)\sin(\pi \wn)\bigg]\, ,
  \end{split}
  \end{equation}
where $C_{\Delta}$ is a normalization constant and we have assumed $T_R=T_L=T$.
The poles of the response function are given by
\begin{equation}\label{QNM_BTZ}
\wn_n(\qn)=\,\pm \qn - i (2n + \Delta),\,\,\,\,\,\,\,n=0,1,2,\, \dots \, .
\end{equation}
The pole-skipping points and critical points of the spectral curve associated with the above response function have been found in \cite{Grozdanov:2019uhi}.   
The nearest critical points to the mode $n=0$ in \eqref{QNM_BTZ} is found to be \cite{Grozdanov:2019uhi}:
\begin{equation}\label{crtical_BTZ}
\qn_c=\pm i,\,\,\,\,w_c=\,-i(1+\Delta).
\end{equation}
On the other hand the radius of convergence of $\wn(\qn)$ expanded around any value of $\qn$  is determined by the closest singularity of $\wn(\qn)$ to that point in the complex $\qn$-plane~\cite{Heller:2020hnq}. 
Since the dispersion relation~\eqref{QNM_BTZ} is an \textit{entire function}, it cannot be singular at the critical points~\eqref{crtical_BTZ}. More precisely,  the dispersion relation~\eqref{crtical_BTZ} does not have any singular point. Thus the radius of convergence of the derivative expansion about the QNMs in the BTZ case is infinite.\footnote{We thank Alexandre Serantes for pointing this out.} 
In summary, we do not expect any relation between pole-skipping and the radius of convergence in the BTZ case. In contrast to BTZ, in $AdS_{d+1}$ with $d>2$,  the dispersion relations have square-root singularities leading to finite radius of convergence about the QNMs, as discussed in last sections.

\section{Discussion}
\label{sec:discussion}
In this paper, we have computed the pole-skipping points and critical points of scalar (spin-0), vector (spin-1), and tensor (spin-2) perturbations with different masses, see sections~\ref{sec:dispersion_constraint} and~\ref{sec:gapped}, respectively. First, this is holographically dual to pole-skipping points in the correlation functions of scalar, vector, and tensor operators with different operator dimensions. Second, it yields the radius of convergence of a low-energy expansion in small  momenta of {\it gapped, non-hydrodynamic} modes. The latter expansion is analogous to the momentum space  expansion of {\it gapless, hydrodynamic} modes. Our analysis yields four main results: 
	\begin{enumerate}
		\item First, the number of pole-skipping points lying on the dispersion relation of the $n^{th}$ QNM decreases with the mode level $n$, as indicated in figure~\ref{fig:dispersion}. The pole-skipping points, i.e.~the dots in figure~\ref{fig:dispersion}, restrict the dispersion relation of each QNM with a given color (red, green and purple, ...) to run through the set of pole-skipping points of the corresponding color. From figure~\ref{fig:dispersion} it is also obvious that the $n^{th}$ QNM runs through the $n^{th}$ pole-skipping point closest to the imaginary momentum axis. That is the pole-skipping point with $n^{th}$ smallest imaginary momentum, $\text{Im}\, \mathfrak{q}$ at a given value of $\text{Im}\,\wn$. For example, the lowest-lying mode with $n=1$ indicated by the red trajectory, runs through those pole-skipping points which at each value of $\text{Im}\,\mathfrak{q}$ are the ($1^{st}$) closest to the $\text{Im}\,\mathfrak{q}$-axis in figure~\ref{fig:dispersion}. 
		\item Second, the pole-skipping points provide an upper bound on the values of the critical points, see eq.~\eqref{eq:upper_bound}. This bound implies that the magnitude of the critical momentum, $|\mathfrak{q}_c^{(1)}|$, of the derivative expansion, $\mathfrak{w}(\mathfrak{q})$, around the lowest gapped quasinormal frequency is bounded from above by the values of the pole-skipping momenta, as illustrated in figure~\ref{fig:Scalar_cross}. 
		This establishes a relation between pole-skipping points and critical points which could not be observed before at fixed values of operator dimensions $\Delta=3,\, 4$~\cite{Grozdanov:2019uhi}. 
		\item Our third result is that for particular values of the operator dimension, $\Delta\approx3.1$, $4.3$, and $\Delta=\Delta_{\text{t}}$, the convergence radius is equal to the momentum of the pole-skipping point, as seen from figure~\ref{fig:Scalar_cross} \footnote{It should be emphasized that the analytic structure of the spectral curve in holography is not known in general. From the known examples, we think that there should not be any non-analyticity in the spectral curve of holographic models. Following this logic and by use of the implicit function theorem, in order to find the radius of convergence,  we focused on critical points of the spectral curve instead of explicitly finding the singular points of dispersion relations. However, we saw that in the BTZ case this simplification failed to work. This is due to the fact that in that case the dispersion relations are exact and non-singular. Thus the main question which remains to be answered is that of the structure of spectral curves in holography~\cite{Heller:2020hnq}.}.  
This indicates that at particular values of $\Delta$ the critical points may be computed by the relatively simple computation for pole-skipping points. This creates a possibility for finding analytic solutions (as opposed to numerical results) for critical points. 
		\item There exists a transition between two distinct types of critical points at a particular conformal dimension $\Delta=\Delta_{\text{t}}$. 
This classification results from the two distinct collision behaviors of the relevant QNMs determining a given critical point. We refer to the first type of QNM behavior at $\Delta< \Delta_\text{t}$ as {\it lowest-level-degeneracy} behavior, illustrated in figure~\ref{fig:Scalar_cross_before}. The behavior of QNMs at $\Delta > \Delta_\text{t}$ is coined {\it level-crossing}, as illustrated in figure~\ref{fig:Scalar_cross_beyond}. This behavior is further detailed in appendix~\ref{sec:Critical_points}.  
	\end{enumerate}
All four results are valid for all operator dimensions we considered, i.e.~in the regimes of level-crossing as well as the regime of lowest-level-degeneracy. These results are also valid  for fields with distinct spins under spatial rotations, i.e.~for massive scalar, massive vector, and massive tensor fluctuations, as discussed in section~\ref{sec:vectorTensor}.   

Figure~\ref{fig:circle} reveals two more features. Critical points above $\Delta_\text{t}$ have complex-valued squared critical momentum, $\mathfrak{q}_c^2$, while for $\Delta<\Delta_\text{t}$ the squared critical momentum is purely imaginary. Figure~\ref{fig:circle} clearly shows that at $\Delta_\text{t}$ a discontinuity occurs: the critical points with purely imaginary momentum instantly jump to critical points with nonzero momentum when $\Delta$ is increased infinitesimally above $\Delta_\text{t}$. 
At $\Delta=\Delta_\text{t}$, the bound~\eqref{eq:upper_bound} from result ``2.'' is saturated and there is a kink in the curve for, $|\mathfrak{q}_c^{(1)}|$ (the radius of convergence about the lowest QNM) as a function of $\Delta$. 
It seems plausible that the kink in this curve is associated with the discontinuity seen in figure~\ref{fig:circle} which we just discussed. 

All the results of this paper are related to gapped QNMs. One naturally may think if analogous results would exist in the case of gapless excitations. In the case of gapless hydrodynamic excitations, the  corresponding pole-skipping points have been discussed in detail in~\cite{Blake:2019otz}. For the shear and diffusion channels, the pole-skipping points lie on the dispersion relations in the Im $\wn$ - Re $\qn$ plane where Im $\wn=-i \,n$. For the sound channel, however, there is not any pole-skipping point at Im $\wn=-i \,n$. Thus, let us focus on the shear and diffusion channels, corresponding to the spin-1 hydrodynamic excitations, namely momentum diffusion and charge diffusion modes. In contrast to the gapped excitations discussed in this work, the pole-skipping in the spin-1 hydrodynamic excitations occurs at real momenta, while the critical points of the dispersion relations have complex momentum~\cite{Grozdanov:2019kge}. It is clear that the latter two sets of points do not lie in the same plane. Then it would be natural not to find any relation analogous to that  of gapped excitations (figure~\ref{fig:Scalar_cross})  for these cases.

Here we have focused on the Schwarzschild background dual to uncharged thermal states. However, we expect our results to hold for other theories and other background metrics as well. As an extension of our analysis one may consider the QNMs of the Reissner-Nordstr\"om~\cite{Abbasi:2020ykq,Jansen:2020hfd}, the rotating~\cite{Garbiso:2020puw} or charged magnetic~\cite{Ammon:2017ded,Ammon:2020rvg,Ammon:2016fru,Grieninger:2017jxz} black branes. It would be interesting to repeat the above computation in theories with gapped half integer spin perturbations corresponding to introducing fermionic operators~\cite{Liu:2009dm,Ammon:2010pg,Ceplak:2019ymw}. 

It is urgently desirable to find analytic solutions for critical points. One way may be to use simpler pole-skipping equations  of motion at the saturation values of $\Delta$ where pole-skipping points coincide with critical points, as we had indicated before. One may also consider different systems, i.e.~different gravitational actions, different black brane solutions, possibly in the extremal limit. The study of $AdS_4$ in~\cite{Grozdanov:2020koi} has shown that the analysis of pole-skipping points is significantly simplified in the lower dimension compared to $AdS_5$. 
In the same vain, it would be interesting to further investigate the saturation values of $\Delta$, perhaps through studying some models which are analytically under control, such as the SYK model~\cite{Choi:2020tdj}. 

Through the connection of the trajectories of hydrodynamic and non-hydrodynamic modes, it is clear that these two types of modes are related in the complexified momentum space.  
Lessons for gapped modes have been learned recently from gapless modes~\cite{Withers:2018srf}. Turning this logic around, the relations we derived for gapped modes in this present work may help to understand gapless modes better.

Pole-skipping points in gapless spin-0 (sound) QNMs have been related to the quantum Lyapunov exponent, $\lambda$, and butterfly velocity, $v_B$, describing the spreading quantum chaos across a system~\cite{Maldacena:2015waa}. While we have derived relations for gapped modes, our results may have implications for quantum chaos and bounds on $\lambda$ and $v_B$. In~\cite{Grozdanov:2019uhi} it was shown that in the case of a Schwarzschild black brane in Einstein gravity the chaos values $\lambda$ and $v_B$ are also encoded in the critical points of the gapless spin-2 excitations of the metric. This may provide a connection between our results and quantum chaos. As a second approach, one may consider an infinitesimal change of the operator dimension of the metric. Then one may repeat our analysis for the spin-0 (sound) QNMs including the now infinitesimally gapped (formerly hydrodynamic sound) modes.  

In \cite{Blake:2019otz,Grozdanov:2019uhi}, the constraints imposed on hydrodynamic modes from pole-skipping points at real momenta have been discussed. It would be interesting to investigate whether the pole-skipping points at imaginary momenta constrain hydrodynamic-dispersion relations, too. 

Finally, it is highly desirable to rigorously prove all of the relations which we have stated in this work based on numerical observations. 
Analytic proofs will allow us to apply our results in practice.  As one example, once one has an analytic proof for the bound discussion in section~\ref{bound}, then it may be possible to find analytic constraints on the response functions. 
In the case of a scalar field theory in the Schwarzschild background,  
this may not be feasible. 
But for solvable models, like that of reference~\cite{Choi:2020tdj}, there would be a better chance for finding relations between pole-skipping and critical points. We aim to check all the observations of the manuscript in that model in the future.   

\acknowledgments
We would like to thank Ali Davody for helpful discussions, and Sa{\v s}o Grozdanov for comments on a draft of this paper. We would like to thank Szabolcs Horvát for providing his excellent MaTeX package. 
This work was supported, in part, by the U.S.~Department of Energy grant DE-SC-0012447, and by grant number 561119208 ``Double First Class'' start-up funding of Lanzhou University, China.

\appendix

\section{Near-horizon expansions \& vector pole-skipping points}
\label{N_H_expansion}
\subsection{Scalar perturbations}
\label{}
The coefficients $M_{rs}$ corresponding to the first four near horizon equations are given in table~\ref{tab:table_near_horizon}. These coefficients are introduced in section~\ref{sec:dispersion_constraint}.
\begin{table}[h]
	\label{table one}
	\centering
	\begin{tabular}[h]{|c|c|}
		\hline
		\hline
		&\\
		$M_{11}$&	$m^2 +4 \qn^2+6 i \wn$\\
				&\\
			\hline             
							&\\
		$M_{21}$&	$3m^2 +4( \qn^2+3i \wn)$\\
		$M_{22}$&	$\pi T(-20+m^2+4\qn^2+18i \wn)$\\
						&\\
				\hline
				&\\
				$M_{31}$&	$3(m^2 +2 i \wn)$\\
				$M_{32}$&	$\pi T(-30+3m^2+4\qn^2+24i \wn)$\\
			$M_{33}$	&$(\pi T)^2(-60+m^2+4\qn^2+30i \wn)$\\
				&\\
				\hline	
					&\\
				$M_{41}$&	$m^2 $\\
				$M_{42}$&	$\pi T(-20+3m^2+10i \wn)$\\
						$M_{43}$&	$(\pi T)^2(3m^2+4(-20+\qn^2+9i\wn))$\\
				$M_{44}$	&$(\pi T)^3(-120+m^2+4\qn^2+42i \wn)$\\
				&\\
				\hline			
		\hline
	\end{tabular}
		\caption{ Examples of coefficients for the equations of motion for the scalar bulk field expanded near the horizon.  
	\label{tab:table_near_horizon}
	}
\end{table}
%
\subsection{Vector perturbations}
\label{}
In analogy to the calculation of scalar perturbation pole-skipping points in section~\ref{sec:dispersion_constraint}, here we sketch the near-horizon calculation for the vector perturbations discussed in section~\ref{sec:vectorTensor}. The transverse vector field is expanded as follows
\begin{equation}\label{}
A_x(r)=\sum_{n=0}^{\infty}a_{n}(r-r_h)^n=\,a_0+(r-r_h)a_1+\cdots \, ,
\end{equation}
and 
\begin{eqnarray}
0&=&N_{11}a_0+4(\pi T)\,(i \wn-1)a_1\, ,\\
0&=&N_{21}a_0+N_{22}a_1+8(\pi T)^2\,(i \wn-2)a_2\, ,\\
0&=&N_{31}a_0+N_{32}a_1+N_{33}a_2+12(\pi T)^3\,(i \wn-3)a_3\, ,\\
0&=&N_{41}a_0+N_{42}a_1+N_{43}a_2+N_{44}a_3+\,16(\pi T)^4\,(i \wn-4)a_4 \, ,
\end{eqnarray}
while the coefficients are given in table~\ref{tab:table_near_horizon_vector}. 
\begin{table}[h]
	\centering
	\begin{tabular}[h]{|c|c|}
		\hline
		\hline
		&\\
		$N_{11}$&	$m^2 +4 \qn^2+2 i \wn$\\
		&\\
		\hline             
		&\\
		$N_{21}$&	$2(m^2+ i \wn)$\\
		$N_{22}$&	$\pi T(-8+m^2+4\qn^2+10i \wn)$\\
		&\\
		\hline
		&\\
		$N_{31}$&	$m^2$\\
		$N_{32}$&	$2\pi T(-5+m^2+3i \wn)$\\
		$N_{33}$	&$(\pi T)^2(-28+m^2+4\qn^2+18i \wn)$\\
		&\\
		\hline	
		&\\
		$N_{41}$&	$0 $\\
		$N_{42}$&	$\pi T(-2+m^2)$\\
		$N_{43}$&	$2(\pi T)^2(-14+m^2+5i\wn)$\\
		$N_{44}$	&$(\pi T)^3(-60+m^2+4\qn^2+26i \wn)$\\
		&\\
		\hline			
		\hline
	\end{tabular}
	\caption{ Examples of coefficients for the equations of motion for the vector bulk field expanded near the horizon.  
		\label{tab:table_near_horizon_vector}
	}
\end{table}
%
The pole-skipping points vor the vector operator are then given by 
\begin{equation}
\begin{split}
\wn=-i:\,\,\,\,\,\,\,\qn^*_{1,1}=&\frac{1}{2}\,i\,\sqrt{\Delta^2-4\Delta+5}\, ,\\
\wn=-2i:\,\,\,\,\,\,\,\qn^*_{j,2}=&\frac{1}{2}\,i\,\sqrt{\Delta^2-4\Delta+11+\,(-1)^j2\sqrt{2}\sqrt{\Delta^2-4\Delta+7}},\,\,\,\,j=1,2 \, .
\end{split}
\end{equation}

\section{Two types of critical points}
\label{sec:Critical_points}
In this section we explicitly illustrate the  positioning of the critical points along the dispersion relations for two cases: for $\Delta=4<\Delta_{\text{t}}$ and then for $\Delta=6>\Delta_{\text{t}}$.

\begin{figure}[h]
	\centering
	\includegraphics[width=0.45\textwidth]{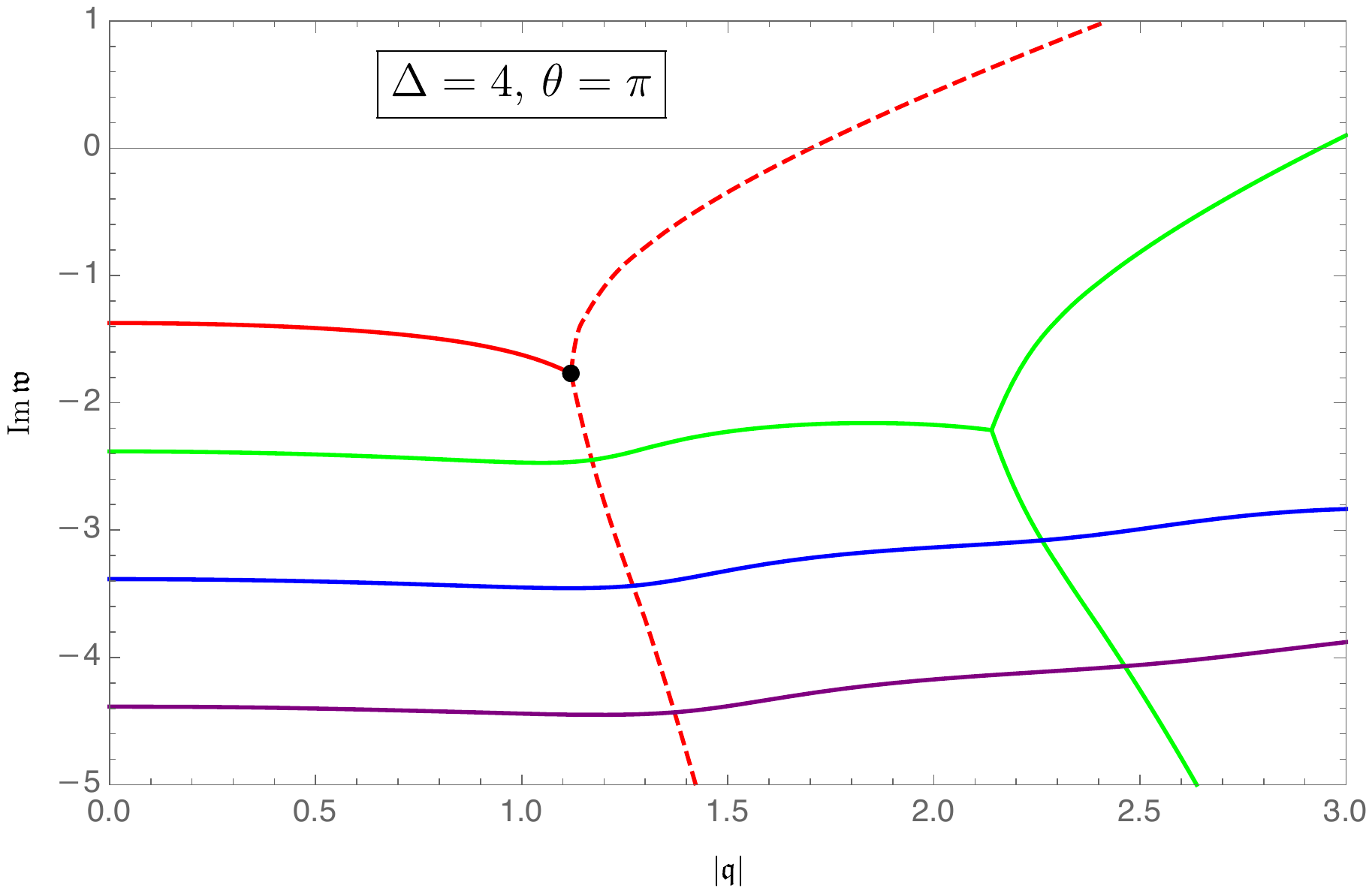}\,\,\,\,\,\,\,\,\includegraphics[width=0.45\textwidth]{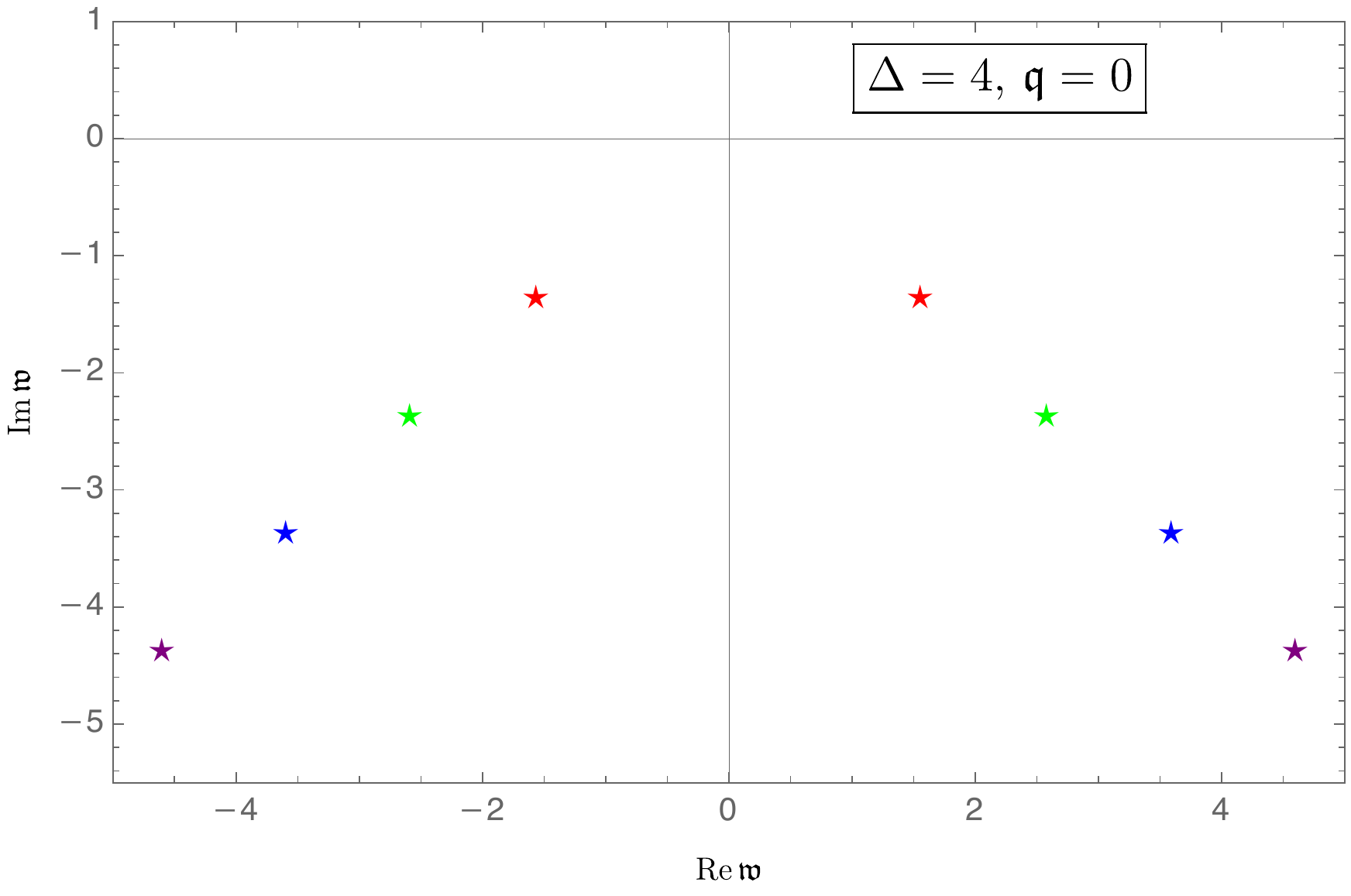} 	
	\caption{Left panel: dispersion relations of the lowest four QNMs at $\Delta=4$ and $\qn^2=|\qn^2|e^{i\theta}$ with $\theta=\pi$. Right panel: spectrum of the lowest four QNMs at $\Delta=4$ and $\qn=0$. The black dot in the left panel is identical to the black dot located on the negative part of the Re$-\qn^2$ axis in figure~\ref{fig:circle}.}
	\label{fig:Collision_Delta_4}
\end{figure}

In figure~\ref{fig:Collision_Delta_4}, we depict the lowest-level-degeneracy which is the indication for the critical point of the spectral curve when $\Delta<\Delta_{\text{t}}$. We are considering the case $\Delta=4$ in this figure.
\newline
As discussed in the main text, the lowest-level-degeneracy corresponds to the collision of the lowest level QNMs at $\qn^2=-|\qn^2|$ on the Im\,$\wn$ axis.  For this reason and by considering $\wn\equiv\wn(\qn^2)=\,\wn(|\qn^2|e^{i \theta})$, we have plotted Im\,$\wn$ as a function of $|\qn|$ at the constant phase $\theta=\pi$. We find that the point at which the red curve splits into two dashed red curves, namely the black dot,  is exactly the collision point we had found before (see black dot in figure~\ref{fig:circle}):
\begin{equation}
\qn_c^2=-1.25309,\,\,\,\,\,\,\,|\qn_c|=\,1.11942 \, .
\end{equation}

\begin{figure}[h]
	\centering
	\includegraphics[width=0.45\textwidth]{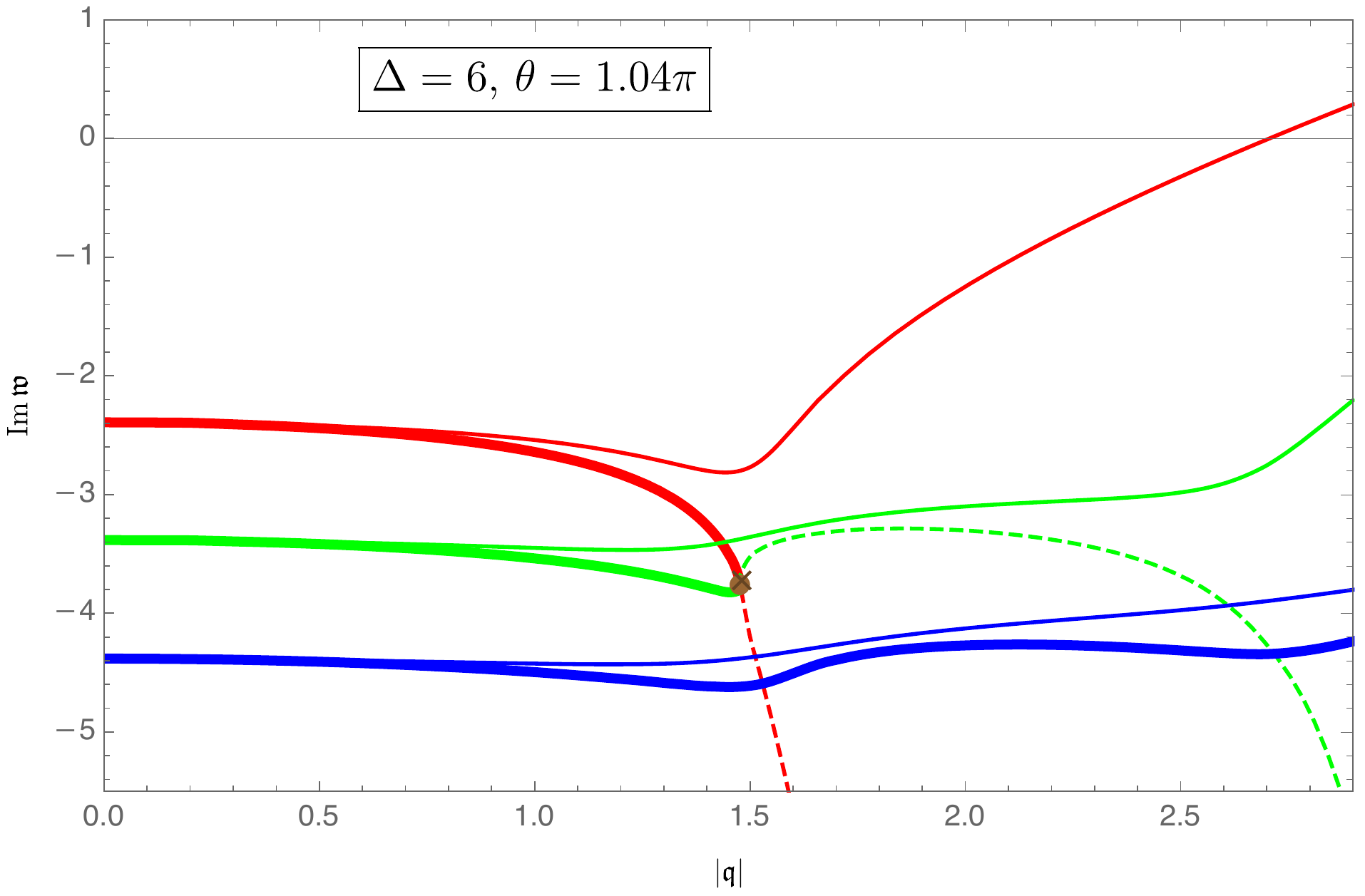}\,\,\,\,\,\,\,\,\includegraphics[width=0.45\textwidth]{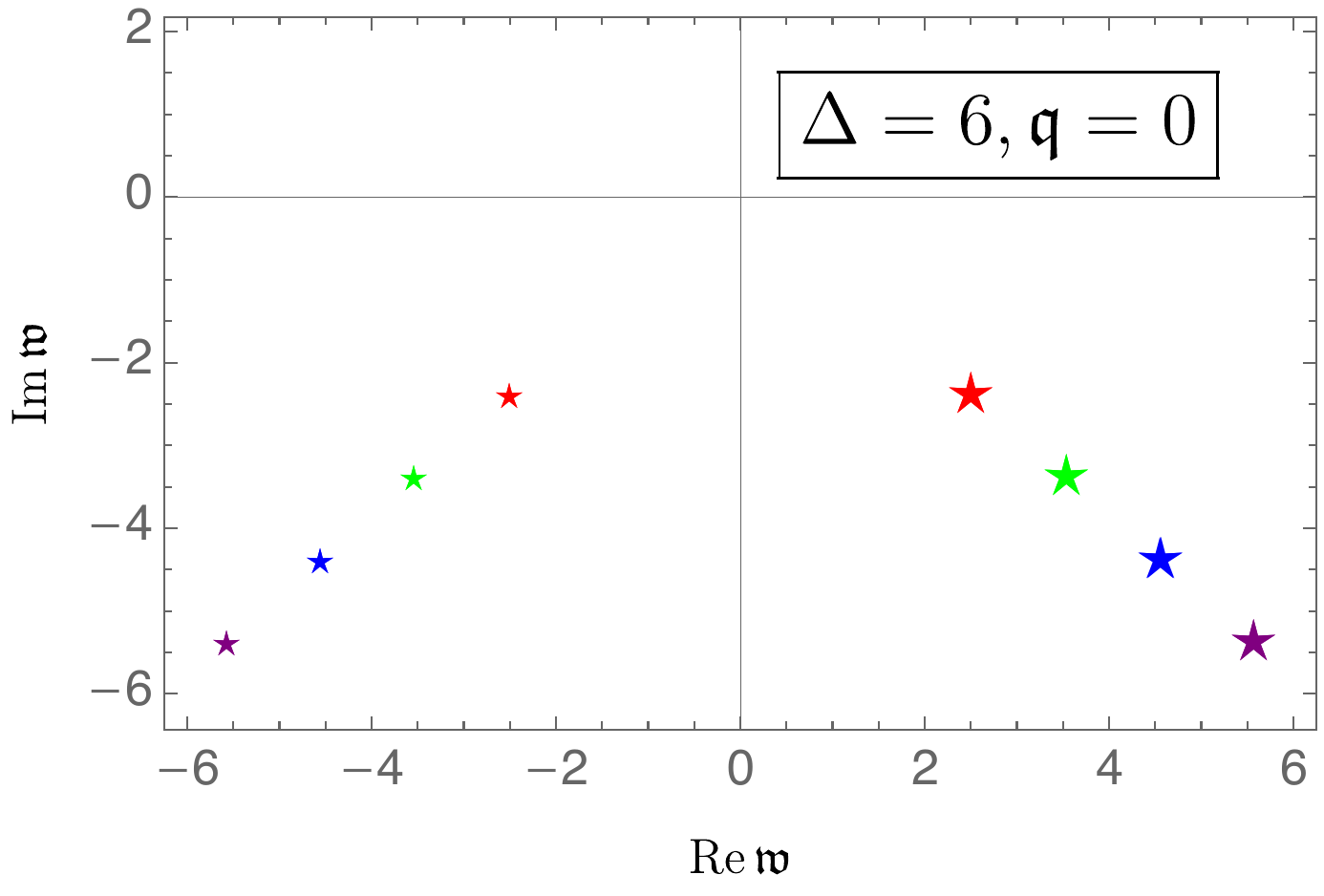} 	
	\caption{Left panel: dispersion relations of the lowest three QNMs at $\Delta=6$ and $\qn^2=|\qn^2|e^{i\theta}$ with $\theta=1.04\pi$. Right panel: spectrum of the lowest three QNMs at $\Delta=6$ and $\qn=0$. The brown cross-dot in the left panel shows exactly the brown cross-dot marked in figure~\ref{fig:circle}. Thick curves in the left panel correspond to the big stars on the right side of the right panel. Dashed curves in the left panel show the dispersion of the modes after the collision.}
	\label{fig:Collision_Delta_6}
\end{figure}

In figure~\ref{fig:Collision_Delta_6}, we have depicted the level-crossing which is the indication for the critical point of the spectral curve when $\Delta>\Delta_{\text{t}}$. We  consider the case $\Delta=6$ in this figure.
\newline
Let us recall that the level-crossing discussed in the text corresponds to the collision of a lowest-lying QNM with the second QNM having Re $\wn$ with the same sign. As shown in figure~\ref{fig:circle}, at $\Delta=6$, this collision occurs at 
\begin{equation}\label{2nd_collision}
\qn_c^2=2.17972\,e^{i (1.04)\pi},\,\,\,\,\,|\qn_c|=1.47638 \, .
\end{equation}
Thus we take $\wn\equiv\wn(\qn^2)=\,\wn\big(|\qn^2|e^{i (1.04)\pi}\big)$ and plot Im\,$\wn-|\qn|$. 
As can be seen in figure~\ref{fig:Collision_Delta_6}, starting from $|\qn|=0$ and keeping $\theta$ constant at $1.04\pi$, the degeneracy between the two lowest-lying modes is split. The splitting becomes wider when $|\qn|$ increases. It means that in the right panel, by taking $\theta=1.04 \pi$, the two red stars have the same $\text{Im}\,\wn$ only at $|\qn|=0$.
For a non-vanishing $|\qn|$ at $\theta=1.04 \pi$, the red star on the right side has a more negative Im $\wn$ than the one in the left side. The former then follows the thick red curve in the left panel.

Also in figure~\ref{fig:Collision_Delta_6} the same behavior is observed for the next higher QNMs with mode number $n=2$. It is indicated by the big green star on the right side of the right panel. Again, keeping fixed $\theta=1.04 \pi$, when $|\qn|$ increases, the green star on the right side follows the thick green curve in the left panel.
At some point, marked by a brown cross-dot, the red and green thick curves collide with each other. This point is exactly \eqref{2nd_collision}. 
The collision between two curves with different colors is the manifestation of level-crossing.
There is also another level-crossing at the same value of $|\qn|$ which corresponds to the collision of small red and and small green stars on the left side of the right panel. Since this collision occurs at $\theta=2\pi-1.04 \pi=0.96 \pi$, it cannot be observed in the left panel.

%
\bibliographystyle{JHEP}
\bibliography{convergence.v2.bib}

\providecommand{\href}[2]{#2}\begingroup\raggedright\begin{thebibliography}{10}

\bibitem{Hiscock:1985zz}
W.~A. Hiscock and L.~Lindblom, \emph{{Generic instabilities in first-order
  dissipative relativistic fluid theories}},
  \href{https://doi.org/10.1103/PhysRevD.31.725}{\emph{Phys. Rev. D} {\bfseries
  31} (1985) 725}.

\bibitem{Kovtun:2019hdm}
P.~Kovtun, \emph{{First-order relativistic hydrodynamics is stable}},
  \href{https://doi.org/10.1007/JHEP10(2019)034}{\emph{JHEP} {\bfseries 10}
  (2019) 034} [\href{https://arxiv.org/abs/1907.08191}{{\ttfamily
  1907.08191}}].

\bibitem{Hoult:2020eho}
R.~E. Hoult and P.~Kovtun, \emph{{Stable and causal relativistic Navier-Stokes
  equations}}, \href{https://doi.org/10.1007/JHEP06(2020)067}{\emph{JHEP}
  {\bfseries 06} (2020) 067}
  [\href{https://arxiv.org/abs/2004.04102}{{\ttfamily 2004.04102}}].

\bibitem{Taghinavaz:2020axp}
F.~Taghinavaz, \emph{{Causality and Stability Conditions of a Conformal Charged
  Fluid}}, \href{https://doi.org/10.1007/JHEP08(2020)119}{\emph{JHEP}
  {\bfseries 08} (2020) 119}
  [\href{https://arxiv.org/abs/2004.01897}{{\ttfamily 2004.01897}}].

\bibitem{Bemfica:2017wps}
F.~S. Bemfica, M.~M. Disconzi and J.~Noronha, \emph{{Causality and existence of
  solutions of relativistic viscous fluid dynamics with gravity}},
  \href{https://doi.org/10.1103/PhysRevD.98.104064}{\emph{Phys. Rev. D}
  {\bfseries 98} (2018) 104064}
  [\href{https://arxiv.org/abs/1708.06255}{{\ttfamily 1708.06255}}].

\bibitem{Bemfica:2019knx}
F.~S. Bemfica, M.~M. Disconzi and J.~Noronha, \emph{{Nonlinear Causality of
  General First-Order Relativistic Viscous Hydrodynamics}},
  \href{https://doi.org/10.1103/PhysRevD.100.104020}{\emph{Phys. Rev. D}
  {\bfseries 100} (2019) 104020}
  [\href{https://arxiv.org/abs/1907.12695}{{\ttfamily 1907.12695}}].

\bibitem{Bemfica:2020zjp}
F.~S. Bemfica, M.~M. Disconzi and J.~Noronha, \emph{{General-Relativistic
  Viscous Fluid Dynamics}},  \href{https://arxiv.org/abs/2009.11388}{{\ttfamily
  2009.11388}}.

\bibitem{Heller:2013fn}
M.~P. Heller, R.~A. Janik and P.~Witaszczyk, \emph{{Hydrodynamic Gradient
  Expansion in Gauge Theory Plasmas}},
  \href{https://doi.org/10.1103/PhysRevLett.110.211602}{\emph{Phys. Rev. Lett.}
  {\bfseries 110} (2013) 211602}
  [\href{https://arxiv.org/abs/1302.0697}{{\ttfamily 1302.0697}}].

\bibitem{Heller:2015dha}
M.~P. Heller and M.~Spalinski, \emph{{Hydrodynamics Beyond the Gradient
  Expansion: Resurgence and Resummation}},
  \href{https://doi.org/10.1103/PhysRevLett.115.072501}{\emph{Phys. Rev. Lett.}
  {\bfseries 115} (2015) 072501}
  [\href{https://arxiv.org/abs/1503.07514}{{\ttfamily 1503.07514}}].

\bibitem{Heller:2020uuy}
M.~P. Heller, A.~Serantes, M.~Spali\'nski, V.~Svensson and B.~Withers,
  \emph{{The hydrodynamic gradient expansion in linear response theory}},
  \href{https://arxiv.org/abs/2007.05524}{{\ttfamily 2007.05524}}.

\bibitem{Withers:2018srf}
B.~Withers, \emph{{Short-lived modes from hydrodynamic dispersion relations}},
  \href{https://doi.org/10.1007/JHEP06(2018)059}{\emph{JHEP} {\bfseries 06}
  (2018) 059} [\href{https://arxiv.org/abs/1803.08058}{{\ttfamily
  1803.08058}}].

\bibitem{Grozdanov:2019kge}
S.~Grozdanov, P.~K. Kovtun, A.~O. Starinets and P.~Tadi\'c, \emph{{Convergence
  of the Gradient Expansion in Hydrodynamics}},
  \href{https://doi.org/10.1103/PhysRevLett.122.251601}{\emph{Phys. Rev. Lett.}
  {\bfseries 122} (2019) 251601}
  [\href{https://arxiv.org/abs/1904.01018}{{\ttfamily 1904.01018}}].

\bibitem{Amado:2008ji}
I.~Amado, C.~Hoyos-Badajoz, K.~Landsteiner and S.~Montero, \emph{{Hydrodynamics
  and beyond in the strongly coupled N=4 plasma}},
  \href{https://doi.org/10.1088/1126-6708/2008/07/133}{\emph{JHEP} {\bfseries
  07} (2008) 133} [\href{https://arxiv.org/abs/0805.2570}{{\ttfamily
  0805.2570}}].

\bibitem{Kaminski:2009ce}
M.~Kaminski, K.~Landsteiner, F.~Pena-Benitez, J.~Erdmenger, C.~Greubel and
  P.~Kerner, \emph{{Quasinormal modes of massive charged flavor branes}},
  \href{https://doi.org/10.1007/JHEP03(2010)117}{\emph{JHEP} {\bfseries 03}
  (2010) 117} [\href{https://arxiv.org/abs/0911.3544}{{\ttfamily 0911.3544}}].

\bibitem{Janiszewski:2015ura}
S.~Janiszewski and M.~Kaminski, \emph{{Quasinormal modes of magnetic and
  electric black branes versus far from equilibrium anisotropic fluids}},
  \href{https://doi.org/10.1103/PhysRevD.93.025006}{\emph{Phys. Rev. D}
  {\bfseries 93} (2016) 025006}
  [\href{https://arxiv.org/abs/1508.06993}{{\ttfamily 1508.06993}}].

\bibitem{Blake:2017ris}
M.~Blake, H.~Lee and H.~Liu, \emph{{A quantum hydrodynamical description for
  scrambling and many-body chaos}},
  \href{https://doi.org/10.1007/JHEP10(2018)127}{\emph{JHEP} {\bfseries 10}
  (2018) 127} [\href{https://arxiv.org/abs/1801.00010}{{\ttfamily
  1801.00010}}].

\bibitem{Blake:2019otz}
M.~Blake, R.~A. Davison and D.~Vegh, \emph{{Horizon constraints on holographic
  Green\textquoteright{}s functions}},
  \href{https://doi.org/10.1007/JHEP01(2020)077}{\emph{JHEP} {\bfseries 01}
  (2020) 077} [\href{https://arxiv.org/abs/1904.12883}{{\ttfamily
  1904.12883}}].

\bibitem{Grozdanov:2019uhi}
S.~Grozdanov, P.~K. Kovtun, A.~O. Starinets and P.~Tadi\'c, \emph{{The complex
  life of hydrodynamic modes}},
  \href{https://doi.org/10.1007/JHEP11(2019)097}{\emph{JHEP} {\bfseries 11}
  (2019) 097} [\href{https://arxiv.org/abs/1904.12862}{{\ttfamily
  1904.12862}}].

\bibitem{Grozdanov:2017ajz}
S.~Grozdanov, K.~Schalm and V.~Scopelliti, \emph{{Black hole scrambling from
  hydrodynamics}},
  \href{https://doi.org/10.1103/PhysRevLett.120.231601}{\emph{Phys. Rev. Lett.}
  {\bfseries 120} (2018) 231601}
  [\href{https://arxiv.org/abs/1710.00921}{{\ttfamily 1710.00921}}].

\bibitem{Blake:2018leo}
M.~Blake, R.~A. Davison, S.~Grozdanov and H.~Liu, \emph{{Many-body chaos and
  energy dynamics in holography}},
  \href{https://doi.org/10.1007/JHEP10(2018)035}{\emph{JHEP} {\bfseries 10}
  (2018) 035} [\href{https://arxiv.org/abs/1809.01169}{{\ttfamily
  1809.01169}}].

\bibitem{Grozdanov:2018kkt}
S.~Grozdanov, \emph{{On the connection between hydrodynamics and quantum chaos
  in holographic theories with stringy corrections}},
  \href{https://doi.org/10.1007/JHEP01(2019)048}{\emph{JHEP} {\bfseries 01}
  (2019) 048} [\href{https://arxiv.org/abs/1811.09641}{{\ttfamily
  1811.09641}}].

\bibitem{Wu:2019esr}
X.~Wu, \emph{{Higher curvature corrections to pole-skipping}},
  \href{https://doi.org/10.1007/JHEP12(2019)140}{\emph{JHEP} {\bfseries 12}
  (2019) 140} [\href{https://arxiv.org/abs/1909.10223}{{\ttfamily
  1909.10223}}].

\bibitem{Ahn:2019rnq}
Y.~Ahn, V.~Jahnke, H.-S. Jeong and K.-Y. Kim, \emph{{Scrambling in Hyperbolic
  Black Holes: shock waves and pole-skipping}},
  \href{https://doi.org/10.1007/JHEP10(2019)257}{\emph{JHEP} {\bfseries 10}
  (2019) 257} [\href{https://arxiv.org/abs/1907.08030}{{\ttfamily
  1907.08030}}].

\bibitem{Li:2019bgc}
W.~Li, S.~Lin and J.~Mei, \emph{{Thermal diffusion and quantum chaos in neutral
  magnetized plasma}},
  \href{https://doi.org/10.1103/PhysRevD.100.046012}{\emph{Phys. Rev. D}
  {\bfseries 100} (2019) 046012}
  [\href{https://arxiv.org/abs/1905.07684}{{\ttfamily 1905.07684}}].

\bibitem{Ceplak:2019ymw}
N.~Ceplak, K.~Ramdial and D.~Vegh, \emph{{Fermionic pole-skipping in
  holography}}, \href{https://doi.org/10.1007/JHEP07(2020)203}{\emph{JHEP}
  {\bfseries 07} (2020) 203}
  [\href{https://arxiv.org/abs/1910.02975}{{\ttfamily 1910.02975}}].

\bibitem{Das:2019tga}
S.~Das, B.~Ezhuthachan and A.~Kundu, \emph{{Real time dynamics from low point
  correlators in 2d BCFT}},
  \href{https://doi.org/10.1007/JHEP12(2019)141}{\emph{JHEP} {\bfseries 12}
  (2019) 141} [\href{https://arxiv.org/abs/1907.08763}{{\ttfamily
  1907.08763}}].

\bibitem{Abbasi:2019rhy}
N.~Abbasi and J.~Tabatabaei, \emph{{Quantum chaos, pole-skipping and
  hydrodynamics in a holographic system with chiral anomaly}},
  \href{https://doi.org/10.1007/JHEP03(2020)050}{\emph{JHEP} {\bfseries 03}
  (2020) 050} [\href{https://arxiv.org/abs/1910.13696}{{\ttfamily
  1910.13696}}].

\bibitem{Liu:2020yaf}
Y.~Liu and A.~Raju, \emph{{Quantum Chaos in Topologically Massive Gravity}},
  \href{https://doi.org/10.1007/JHEP12(2020)027}{\emph{JHEP} {\bfseries 12}
  (2020) 027} [\href{https://arxiv.org/abs/2005.08508}{{\ttfamily
  2005.08508}}].

\bibitem{Ahn:2020bks}
Y.~Ahn, V.~Jahnke, H.-S. Jeong, K.-Y. Kim, K.-S. Lee and M.~Nishida,
  \emph{{Pole-skipping of scalar and vector fields in hyperbolic space:
  conformal blocks and holography}},
  \href{https://doi.org/10.1007/JHEP09(2020)111}{\emph{JHEP} {\bfseries 09}
  (2020) 111} [\href{https://arxiv.org/abs/2006.00974}{{\ttfamily
  2006.00974}}].

\bibitem{Moitra:2020dal}
U.~Moitra, S.~K. Sake and S.~P. Trivedi, \emph{{Near-Extremal Fluid
  Mechanics}},  \href{https://arxiv.org/abs/2005.00016}{{\ttfamily
  2005.00016}}.

\bibitem{Abbasi:2020ykq}
N.~Abbasi and S.~Tahery, \emph{{Complexified quasinormal modes and the
  pole-skipping in a holographic system at finite chemical potential}},
  \href{https://doi.org/10.1007/JHEP10(2020)076}{\emph{JHEP} {\bfseries 10}
  (2020) 076} [\href{https://arxiv.org/abs/2007.10024}{{\ttfamily
  2007.10024}}].

\bibitem{Jansen:2020hfd}
A.~Jansen and C.~Pantelidou, \emph{{Quasinormal modes in charged fluids at
  complex momentum}},
  \href{https://doi.org/10.1007/JHEP10(2020)121}{\emph{JHEP} {\bfseries 10}
  (2020) 121} [\href{https://arxiv.org/abs/2007.14418}{{\ttfamily
  2007.14418}}].

\bibitem{Grozdanov:2020koi}
S.~Grozdanov, \emph{{Bounds on transport from univalence and pole-skipping}},
  \href{https://arxiv.org/abs/2008.00888}{{\ttfamily 2008.00888}}.

\bibitem{Ramirez:2020qer}
D.~M. Ramirez, \emph{{Chaos and pole skipping in CFT$_2$}},
  \href{https://arxiv.org/abs/2009.00500}{{\ttfamily 2009.00500}}.

\bibitem{Choi:2020tdj}
C.~Choi, M.~Mezei and G.~S\'arosi, \emph{{Pole skipping away from maximal
  chaos}},  \href{https://arxiv.org/abs/2010.08558}{{\ttfamily 2010.08558}}.

\bibitem{Ahn:2020baf}
Y.~Ahn, V.~Jahnke, H.-S. Jeong, K.-Y. Kim, K.-S. Lee and M.~Nishida,
  \emph{{Classifying pole-skipping points}},
  \href{https://arxiv.org/abs/2010.16166}{{\ttfamily 2010.16166}}.

\bibitem{Natsuume:2020snz}
M.~Natsuume and T.~Okamura, \emph{{Pole-skipping and zero temperature}},
  \href{https://arxiv.org/abs/2011.10093}{{\ttfamily 2011.10093}}.

\bibitem{Arean:2020eus}
D.~Arean, R.~A. Davison, B.~Gout\'eraux and K.~Suzuki, \emph{{Hydrodynamic
  diffusion and its breakdown near AdS$_2$ fixed points}},
  \href{https://arxiv.org/abs/2011.12301}{{\ttfamily 2011.12301}}.

\bibitem{Kim:2020url}
K.-Y. Kim, K.-S. Lee and M.~Nishida, \emph{{Holographic scalar and vector
  exchange in OTOCs and pole-skipping phenomena}},
  \href{https://arxiv.org/abs/2011.13716}{{\ttfamily 2011.13716}}.

\bibitem{Sil:2020jhr}
K.~Sil, \emph{{Pole skipping and chaos in anisotropic plasma: a holographic
  study}},  \href{https://arxiv.org/abs/2012.07710}{{\ttfamily 2012.07710}}.

\bibitem{Baggioli:2020loj}
M.~Baggioli, \emph{{How small hydrodynamics can go}},
  \href{https://arxiv.org/abs/2010.05916}{{\ttfamily 2010.05916}}.

\bibitem{Maldacena:2015waa}
J.~Maldacena, S.~H. Shenker and D.~Stanford, \emph{{A bound on chaos}},
  \href{https://doi.org/10.1007/JHEP08(2016)106}{\emph{JHEP} {\bfseries 08}
  (2016) 106} [\href{https://arxiv.org/abs/1503.01409}{{\ttfamily
  1503.01409}}].

\bibitem{Policastro:2002se}
G.~Policastro, D.~T. Son and A.~O. Starinets, \emph{{From AdS / CFT
  correspondence to hydrodynamics}},
  \href{https://doi.org/10.1088/1126-6708/2002/09/043}{\emph{JHEP} {\bfseries
  09} (2002) 043} [\href{https://arxiv.org/abs/hep-th/0205052}{{\ttfamily
  hep-th/0205052}}].

\bibitem{Witten:1998qj}
E.~Witten, \emph{{Anti-de Sitter space and holography}},
  \href{https://doi.org/10.4310/ATMP.1998.v2.n2.a2}{\emph{Adv. Theor. Math.
  Phys.} {\bfseries 2} (1998) 253}
  [\href{https://arxiv.org/abs/hep-th/9802150}{{\ttfamily hep-th/9802150}}].

\bibitem{Nunez:2003eq}
A.~Nunez and A.~O. Starinets, \emph{{AdS / CFT correspondence, quasinormal
  modes, and thermal correlators in N=4 SYM}},
  \href{https://doi.org/10.1103/PhysRevD.67.124013}{\emph{Phys. Rev. D}
  {\bfseries 67} (2003) 124013}
  [\href{https://arxiv.org/abs/hep-th/0302026}{{\ttfamily hep-th/0302026}}].

\bibitem{Heller:2020hnq}
M.~P. Heller, A.~Serantes, M.~Spali\'nski, V.~Svensson and B.~Withers,
  \emph{{Convergence of hydrodynamic modes: insights from kinetic theory and
  holography}},  \href{https://arxiv.org/abs/2012.15393}{{\ttfamily
  2012.15393}}.

\bibitem{Aharony:1999ti}
O.~Aharony, S.~S. Gubser, J.~M. Maldacena, H.~Ooguri and Y.~Oz, \emph{{Large N
  field theories, string theory and gravity}},
  \href{https://doi.org/10.1016/S0370-1573(99)00083-6}{\emph{Phys. Rept.}
  {\bfseries 323} (2000) 183}
  [\href{https://arxiv.org/abs/hep-th/9905111}{{\ttfamily hep-th/9905111}}].

\bibitem{Kabat:2012hp}
D.~Kabat, G.~Lifschytz, S.~Roy and D.~Sarkar, \emph{{Holographic representation
  of bulk fields with spin in AdS/CFT}},
  \href{https://doi.org/10.1103/PhysRevD.86.026004}{\emph{Phys. Rev. D}
  {\bfseries 86} (2012) 026004}
  [\href{https://arxiv.org/abs/1204.0126}{{\ttfamily 1204.0126}}].

\bibitem{Minwalla:1997ka}
S.~Minwalla, \emph{{Restrictions imposed by superconformal invariance on
  quantum field theories}},
  \href{https://doi.org/10.4310/ATMP.1998.v2.n4.a4}{\emph{Adv. Theor. Math.
  Phys.} {\bfseries 2} (1998) 783}
  [\href{https://arxiv.org/abs/hep-th/9712074}{{\ttfamily hep-th/9712074}}].

\bibitem{Benini:2010pr}
F.~Benini, C.~P. Herzog, R.~Rahman and A.~Yarom, \emph{{Gauge gravity duality
  for d-wave superconductors: prospects and challenges}},
  \href{https://doi.org/10.1007/JHEP11(2010)137}{\emph{JHEP} {\bfseries 11}
  (2010) 137} [\href{https://arxiv.org/abs/1007.1981}{{\ttfamily 1007.1981}}].

\bibitem{Son:2002sd}
D.~T. Son and A.~O. Starinets, \emph{{Minkowski space correlators in AdS / CFT
  correspondence: Recipe and applications}},
  \href{https://doi.org/10.1088/1126-6708/2002/09/042}{\emph{JHEP} {\bfseries
  09} (2002) 042} [\href{https://arxiv.org/abs/hep-th/0205051}{{\ttfamily
  hep-th/0205051}}].

\bibitem{Garbiso:2020puw}
M.~Garbiso and M.~Kaminski, \emph{{Hydrodynamics of simply spinning black holes
  \& hydrodynamics for spinning quantum fluids}},
  \href{https://doi.org/10.1007/JHEP12(2020)112}{\emph{JHEP} {\bfseries 12}
  (2020) 112} [\href{https://arxiv.org/abs/2007.04345}{{\ttfamily
  2007.04345}}].

\bibitem{Ammon:2017ded}
M.~Ammon, M.~Kaminski, R.~Koirala, J.~Leiber and J.~Wu, \emph{{Quasinormal
  modes of charged magnetic black branes \textbackslash{}\& chiral magnetic
  transport}}, \href{https://doi.org/10.1007/JHEP04(2017)067}{\emph{JHEP}
  {\bfseries 04} (2017) 067}
  [\href{https://arxiv.org/abs/1701.05565}{{\ttfamily 1701.05565}}].

\bibitem{Ammon:2020rvg}
M.~Ammon, S.~Grieninger, J.~Hernandez, M.~Kaminski, R.~Koirala, J.~Leiber
  et~al., \emph{{Chiral hydrodynamics in strong magnetic fields}},
  \href{https://arxiv.org/abs/2012.09183}{{\ttfamily 2012.09183}}.

\bibitem{Ammon:2016fru}
M.~Ammon, S.~Grieninger, A.~Jimenez-Alba, R.~P. Macedo and L.~Melgar,
  \emph{{Holographic quenches and anomalous transport}}, {\emph{JHEP}
  {\bfseries 09} (2016) 131}
  [\href{https://arxiv.org/abs/1607.06817}{{\ttfamily 1607.06817}}].

\bibitem{Grieninger:2017jxz}
S.~Grieninger, \emph{{Holographic quenches and anomalous transport}},  Master's
  thesis, Jena U., TPI, 2016.

\bibitem{Liu:2009dm}
H.~Liu, J.~McGreevy and D.~Vegh, \emph{{Non-Fermi liquids from holography}},
  \href{https://doi.org/10.1103/PhysRevD.83.065029}{\emph{Phys. Rev. D}
  {\bfseries 83} (2011) 065029}
  [\href{https://arxiv.org/abs/0903.2477}{{\ttfamily 0903.2477}}].

\bibitem{Ammon:2010pg}
M.~Ammon, J.~Erdmenger, M.~Kaminski and A.~O'Bannon, \emph{{Fermionic Operator
  Mixing in Holographic p-wave Superfluids}},
  \href{https://doi.org/10.1007/JHEP05(2010)053}{\emph{JHEP} {\bfseries 05}
  (2010) 053} [\href{https://arxiv.org/abs/1003.1134}{{\ttfamily 1003.1134}}].

\end{thebibliography}\endgroup

%
%

\end{document}